\documentclass[%
 aip,
 amsmath,amssymb,
reprint,%
]{revtex4-1}

\usepackage{graphicx}
\usepackage{dcolumn}
\usepackage{bm}

\usepackage{setspace}
\usepackage[T1]{fontenc}
\usepackage{mathptmx}
\usepackage{etoolbox}
\usepackage{float}
\usepackage{tabularx}
\usepackage{xcolor}

\usepackage{amsmath,amssymb}

\usepackage{appendix}

\begin{document}

\singlespacing

\title{Simulating structured fluids with tensorial viscoelasticity}
\author{Carlos Floyd}
\affiliation{Chicago Center for Theoretical Chemistry, University of Chicago, Chicago, Illinois 60637, USA
}
\affiliation{Department of Chemistry, University of Chicago, Chicago, Illinois 60637, USA
}
\affiliation{James Franck Institute, University of Chicago, Chicago, Illinois 60637, USA
}
\author{Suriyanarayanan Vaikuntanathan}
\email{svaikunt@uchicago.edu}
\affiliation{Chicago Center for Theoretical Chemistry, University of Chicago, Chicago, Illinois 60637, USA
}
\affiliation{Department of Chemistry, University of Chicago, Chicago, Illinois 60637, USA
}
\affiliation{James Franck Institute, University of Chicago, Chicago, Illinois 60637, USA
}

\author{Aaron R.\ Dinner}
\email{dinner@uchicago.edu}
\affiliation{Chicago Center for Theoretical Chemistry, University of Chicago, Chicago, Illinois 60637, USA
}
\affiliation{Department of Chemistry, University of Chicago, Chicago, Illinois 60637, USA
}
\affiliation{James Franck Institute, University of Chicago, Chicago, Illinois 60637, USA
}

\date{\today}

\begin{abstract}
We consider an immersed elastic body that is actively driven through a structured fluid by a motor or an external force.  The behavior of such a system generally cannot be solved analytically, necessitating the use of numerical methods.  However, current numerical methods omit important details of the microscopic structure and dynamics of the fluid, which can modulate the magnitudes and directions of viscoelastic restoring forces. To address this issue, we develop a simulation platform for modeling viscoelastic media with tensorial elasticity.  We build on the lattice Boltzmann algorithm and incorporate viscoelastic forces, elastic immersed objects, a microscopic orientation field, and coupling between viscoelasticity and the orientation field.  We demonstrate our method by characterizing how the viscoelastic restoring force on a driven immersed object depends on various key parameters as well as the tensorial character of the elastic response.  We find that the restoring force depends non-monotonically on the rate of diffusion of the stress and the size of the object.  We further show how the restoring force depends on the relative orientation of the microscopic structure and the pulling direction.  These results imply that accounting for previously neglected physical features, such as stress diffusion and the microscopic orientation field, can improve the realism of viscoelastic simulations.  We discuss possible applications and extensions to the method.
\end{abstract}

\maketitle

\section{Introduction}

When a viscoelastic material responds to to an external mechanical perturbation, elasticity gives rise to restoring forces that push the system back towards its unperturbed configuration, while viscosity dissipates these forces on a material-dependent timescale \cite{phan2013understanding, bird1987dynamics}.  These dynamics can be exploited to engineer metamaterials with new functionalities \cite{dykstra2022extreme} and energy-absorbing media \cite{gutierrez2014engineering, borcherdt2009viscoelastic}.  They also underlie a form of memory:  canonical viscoelastic systems such as structured fluids---dense solutions of interacting particles such as colloids, surfactants, or polymers \cite{witten1990structured}---can adapt their microstructure in response to a shear protocol \cite{paulsen2014multiple,majumdar2018mechanical,scheff2021actin}, in some cases ``learning'' multiple frequencies simultaneously \cite{paulsen2014multiple}.  


In biological contexts, mechanical memory allows cells to adapt their structural responses to forces, enhancing various physiological functions \cite{banerjee2020actin}.  Within cells, the viscoelastic dynamics of structured fluids are also important.  For example, the actin cytoskeleton acts like a viscoelastic medium that governs the motions of vesicles and other objects that are driven through the cytoplasm by molecular motors \cite{fakhri2014high,katrukha2017probing}.  The viscoelasticty of the cytoplasm has even been shown to guide the positioning of the mitotic spindle for cell division \cite{Xiee2115593119}.  Experimental protocols measuring the forces on magnetic particles or objects in optical traps are often used to study the viscoelastic response of the cytoplasm and materials assembled from its constituents \cite{amblard1996magnetic, bausch1999measurement, mas2013quantitative, jun2014calibration}.

If a structured fluid has an anisotropic component, such as a polymeric network like the actin cytoskeleton \cite{fletcher2010cell}, then the viscoelastic forces can have both orientation and timescale dependent characteristics.  
Quantitatively describing these characteristics is important for understanding dynamics in such a medium and interpreting experiments that probe them.   
However, existing theoretical treatments are typically limited to isotropic fluids, neglecting the spatial variation of the microscopic distribution over the solvated molecule's orientations.  Theoretical treatments of anisotropic structured fluids of which we are aware \cite{volkov2007basic,kwon2008microstructurally} consider only the fluid and thus are not sufficient by themselves to model the dynamics of immersed objects such as the mitotic spindle, vesicles, or probe particles mentioned above.

Several mathematical treatments of the stochastic dynamics of a Brownian particle immersed in a viscoelastic medium have been extended to account for confining potentials that bias the particle's motion \cite{dekee2002transport, raikher2013brownian, paul2018free, raikher2013brownian, mankin2013memory}.  However, these studies focus only on the position of the particle itself and do not allow one to spatially resolve the motion of the surrounding medium, which in the cytoplasm for instance can include non-trivial flow patterns \cite{Xiee2115593119}.  To resolve an immersed body together with its surrounding medium, it is necessary to use numerical methods to solve the Navier-Stokes equation.  Although techniques have been introduced to address various aspects of this problem, they have not previously been integrated to treat the response of a structured fluid with a spatially varying tensorial response (e.g., due to colloidal or polymeric orientation) to the movement of an immersed body.  Such a tensorial response can allow for structural memory effects, in which the history of applied stress on, e.g., a polymer network like the actin cytoskeleton creates anisotropic mechanical compliance and even mechanical hysteresis \cite{scheff2021actin}.

In this work, we adapt a tensorial model for viscoelasticity with an explicit microscopic orientation field \cite{kwon2008microstructurally} such that it can be simulated efficiently, and we marry it with the lattice Boltzmann \cite{kruger2017lattice} and immersed boundary \cite{peskin1972flow, peskin2002immersed, feng2004immersed, kruger2014deformability} approaches to create detailed simulations of an elastic body being driven through a structured fluid.  We demonstrate this framework's use for studying physical effects that were previously neglected in hydrodynamic simulations, such as the dependence of viscoelastic return on stress diffusion and the contribution of the underlying orientation field to the elastic response.  This framework can be applied in future work to investigate the interplay between viscoelastic restoring forces, driven immersed objects, and the dynamics of molecular orientations.

\section{Methods}\label{Methods}
In Section \ref{sec:IBLB} we briefly summarize the lattice Boltzmann (LB) and immersed boundary (IB) methods that serve as the foundation of our approach, and we provide further details in Appendix \ref{IBLBApp}.  As a first step toward simulating viscoelastic fluids, we consider in Section \ref{sec:scalar} how to incorporate a scalar model of viscoelasticity into the IB-LB framework.  In other words, here we first assume that elastic effects can be adequately captured by the scalar quantity $C$ representing the elastic modulus.  The methods presented in this Section \ref{sec:scalar} have been developed in previous work, and they form the basis for our treatment of tensorial elasticity in the Section \ref{sec:TVM}.  There, we generalize the model to allow the elastic modulus to be a rank-four tensor $\mathbf{C}(\mathbf{P})$ that depends on the local mean orientation $\mathbf{P}(\mathbf{r})$ of an explicitly tracked polymer field.  For concreteness, in this paper we refer to the viscoelastic medium as being ``polymeric,'' but our methods encompass other structured fluids such as solutions of anisotropic colloids or worm-like micelles.  

\subsection{Immersed boundary-Lattice Boltzmann method}\label{sec:IBLB}

We represent a fluid by its local mass density $\rho(\mathbf{r},t)$ and velocity $\mathbf{v}(\mathbf{r},t)$ which evolve according to 
\begin{equation}
    \partial_t \rho + \partial_i (\rho v_i) = 0, \label{eqDensCons} 
\end{equation}
\begin{equation}
    \rho D_t v_i = - \partial_i p + \partial_j \sigma_{ij}^\text{v} + f_i. \label{eqNS}
\end{equation}
In these equations, $\partial_t$ is a partial derivative with respect to time, $D_t = \partial_t + v_k \partial_k$ is the material derivative, $p$ is the hydrostatic pressure, $\sigma_{ij}^\text{v} = \eta_\text{s} \left(\partial_i v_j + \partial_j v_i \right)$ is the viscous stress tensor, $\eta_\text{s}$ is the fluid's dynamic viscosity, and $f_i$ is a body force density which in later sections encompasses additional aspects of the system's physics. The indices $i$ and $j$ are for Cartesian directions, and repeated indices imply summation.  Equation \ref{eqDensCons} represents conservation of mass, and Equation \ref{eqNS} is the Navier-Stokes equation.  

The standard LB method is a versatile approach to computational fluid dynamics that numerically solves Equations \ref{eqDensCons} and \ref{eqNS}.  For many systems of interest it has practical advantages over alternative methods such as molecular dynamics, lattice gas models, finite element methods, and dissipative particle dynamics \cite{kruger2017lattice}.  These advantages include comparative ease of implementation, high computational efficiency for many systems, flexibility in handling complex boundary geometries and conditions, ability to handle multi-component and multi-phase flows, and a strong physical basis rooted in the Boltzmann equation \cite{he1997theory}.

The LB method represents a fluid on a regular grid with spacing $\Delta x$ in terms of its phase space distribution.  The quantities $h_n(\mathbf{r}, t), \ n = 1,\ldots , N_{LB}$, denote the density of fluid at position $\mathbf{r}$ and time $t$ which has velocity $\mathbf{c}_n$.  The $N_{LB}$ fixed vectors $\{ \mathbf{c}_n \}_{n=1}^{N_{LB}}$ form a discrete set onto which the velocity part of the distribution function is expanded.  The choice of $N_{LB}$ vectors, along with the system dimensionality $d$, define the basic lattice structure of the simulation.  In this paper we take $d = 2$ and $N_{LB} = 9$, the so-called $d2Q9$ lattice.  The macroscopic density $\rho(\mathbf{r}, t)$ and velocity $\mathbf{v}(\mathbf{r}, t)$ are obtained from the distribution functions as 
\begin{eqnarray}
    \rho(\mathbf{r}, t) &=& \sum_{n=1}^{N_{LB}} h_n(\mathbf{r},t), \label{eqrhoDef} \\
    \rho(\mathbf{r}, t)\mathbf{v}(\mathbf{r}, t) &=& \sum_{n=1}^{N_{LB}} \mathbf{c}_n h_n(\mathbf{r},t). \label{eqvDef}
\end{eqnarray}
To simulate the fluid's dynamics, one iterates between collision and streaming steps.  In the collision steps, the distribution functions $h_n(\mathbf{r}, t)$ at each node are relaxed toward their local equilibrium values to represent the redistribution of fluid velocities.  In the streaming steps, the distribution functions are propagated along their corresponding velocity vector $\mathbf{c}_n$ to represent the motion of that fluid population to a neighboring node.  The combined collision and streaming step can be expressed as 
\begin{equation}
    h_n(\mathbf{r} + \mathbf{c}_n \Delta t, t + \Delta t) - h_n(\mathbf{r},t) = \mathcal{C}_n(\mathbf{r},t), \label{eqLB}
\end{equation}
where $\Delta t$ is the simulation timestep  and $\mathcal{C}_n(\mathbf{r},t)$ is the collision operator.  The commonly used Bhatnagar-Gross-Krook form for this operator is 
\begin{equation}
    \mathcal{C}_n(\mathbf{r},t) = -\frac{\Delta t}{\tau}\left(h_n(\mathbf{r},t) - h_n^{\text{eq}}(\mathbf{r},t) \right), \label{eqBGK}
\end{equation}
where $\tau$ is the relaxation time, related to the kinematic viscosity $\nu_\text{s}$ and the speed of sound constant for the lattice $c_\text{s} = \Delta{x} / \sqrt{3} \Delta t$ through 
\begin{equation}
    \nu_\text{s} = c_\text{s}^2\left(\tau - \frac{\Delta t}{2} \right).
\end{equation}  
In Equation \ref{eqBGK}, $h_n^{\text{eq}}(\mathbf{r},t)$ is the local equilibrium distribution, which depends on the macroscopic density and velocity in Equations \ref{eqrhoDef} and \ref{eqvDef}.  Equation \ref{eqBGK} thus couples $h_n(\mathbf{r},t)$ to the remaining $h_m(\mathbf{r},t), \ m \neq n$.  We omit here the expression for $h_n^{\text{eq}}(\mathbf{r},t)$, which can be found for instance in Ref. \citenum{kruger2017lattice}.  It can be shown via Chapman-Enskog analysis that the ``lattice Boltzmann equation," Equation \ref{eqLB}, recovers Equations \ref{eqDensCons} and \ref{eqNS} to second order in the Knudsen number of the fluid.

We note that the LB algorithm introduced here assumes that the fluid is finitely compressible, with an isothermal equation of state for the pressure
\begin{equation}
    p = c_\text{s}^2 \rho.
\end{equation}
Given that the fluid is compressible, there could be a contribution to the viscous stress arising from the divergence of the velocity field in addition to the contribution from the shear viscosity.  This bulk viscosity contribution to the stress is, however, not included as a term in Equation \ref{eqNS}, resulting from the particular choice $\eta_\text{B} = (2/3) \eta_\text{s}$ for the solvent's bulk viscosity coefficient \cite{kruger2017lattice}.  Variations of the LB algorithm are possible which generalize this choice \cite{dellar2001bulk}, but we do not consider these here. 

The LB method can be extended to simulate an elastic body interacting with a fluid environment.  For this, we use the LB method to treat the hydrodynamics of the fluid and the immersed boundary (IB) method to treat the dynamics of the body \cite{peskin1972flow, peskin2002immersed, feng2004immersed, kruger2014deformability}.  In Appendix \ref{IBLBApp}, we review the IB-LB method, and we refer readers to Refs.\ \citenum{kruger2017lattice} and \citenum{kang2011comparative} for detailed descriptions.  The IB method is suited to simulating objects which are deformable and permeable to the fluid, such as an immiscible droplet or vesicle.  To simulate rigid and volume excluding objects such as a bead, alternative numerical methods, like a moving hard-wall boundary, need to be used.  However, the model of viscoelasticity which we present in this work should still be applicable to treating the fluid.  

\subsection{Scalar viscoelastic model}\label{sec:scalar}

Early implementations of scalar viscoelasticity in LB simulations altered the collision operator in Equation \ref{eqBGK} or introduced additional fluid density functions to describe polymer orientations \cite{qian1997lattice, giraud1997lattice, ma2020immersed}.  Both of these approaches require substantial modifications to the basic LB algorithm and do not easily accommodate varying the viscoelastic model, which limits flexibility.  More recent studies have instead favored introducing viscoelastic effects via an additive contribution $\boldsymbol{\sigma}^\text{p}$ to the total stress tensor \cite{ispolatov2002lattice, malaspinas2010lattice, dzanic2022hybrid}.  The divergence of $\boldsymbol{\sigma}^\text{p}$ is a force density $f_i^\text{p} = \partial_j \sigma_{ij}^\text{p}$ which can be included in the force $\mathbf{f}$ appearing in Equation \ref{eqNS}.  We incorporate this force in the LB algorithm using the method introduced in Ref.\ \citenum{guo2002discrete}.  A key benefit of this approach is that it is highly modular, allowing the dynamics of $\boldsymbol{\sigma}^\text{p}$ to be specified independently from the underlying LB algorithm.

\begin{figure}
\begin{center}
\includegraphics[width= \columnwidth]{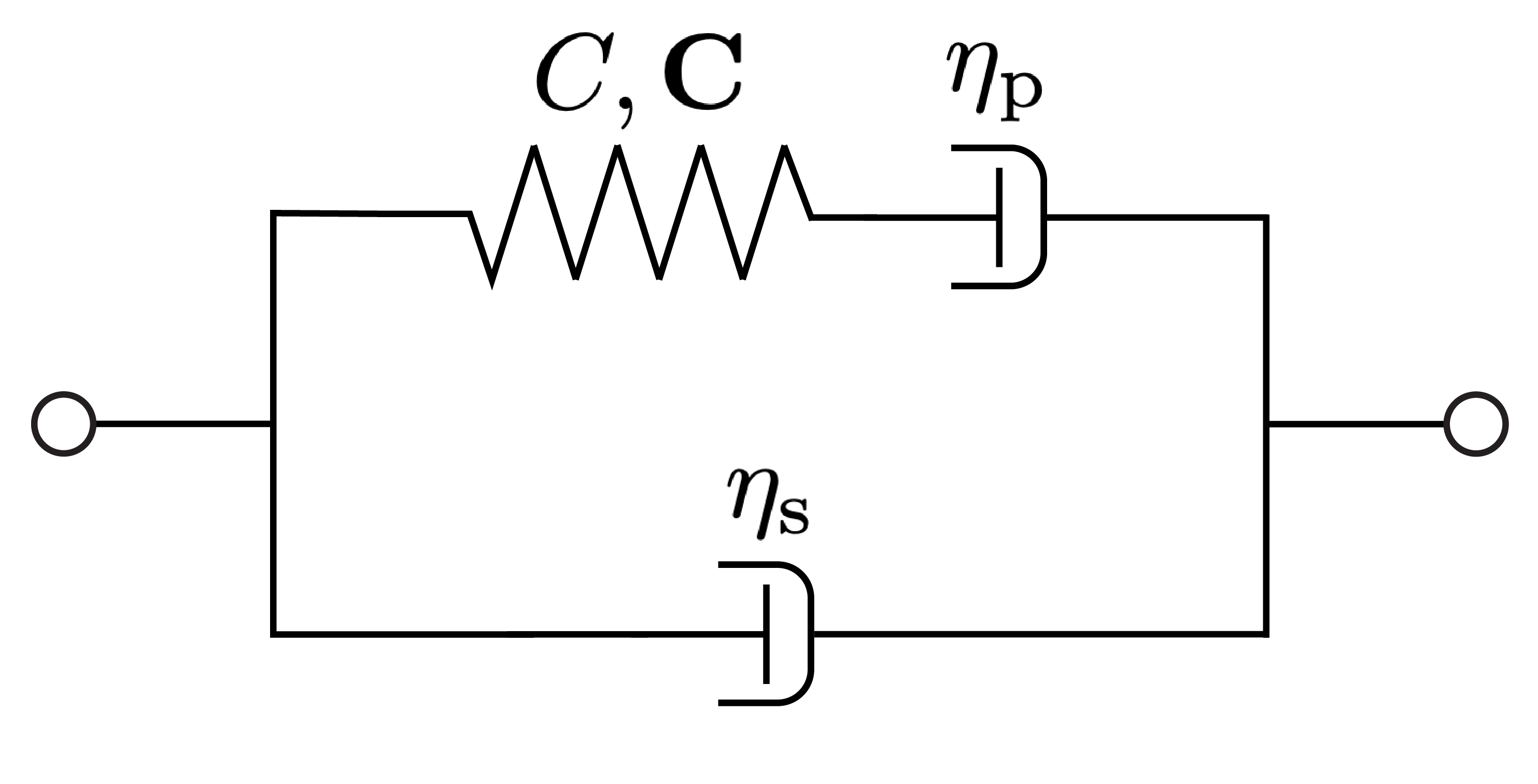}
\caption{Circuit diagram of the viscoelastic model used for simulation. The polymeric elasticity is treated either through a scalar elasticity modulus ($C$) or an elasticity tensor ($\mathbf{C}$). The polymeric elasticity acts in series with the polymeric viscosity ($\eta_\text{p}$), and, together, they act in parallel with the solvent viscosity ($\eta_\text{s}$).    }
\label{Jeffreys}
\end{center}
\end{figure}

In the Jeffreys fluid model, which we use here, the solvent and viscoelastic systems act in parallel, which implies that their corresponding forces are added together in Equation \ref{eqNS} \cite{bird1987dynamics}. The viscoelastic part is further treated using the Maxwell model, in which the elastic and viscous contributions act in series (Figure \ref{Jeffreys}).   The elastic contribution to the viscoelastic stress is then the same as the viscous contribution, which we denote $\boldsymbol{\sigma}^\text{p}$.  From linear viscoelascity theory \cite{phan2013understanding, bird1987dynamics, larson2013constitutive},
\begin{equation}\label{eqlinve}
    \boldsymbol{\sigma}^\text{p} =  C\boldsymbol{\Lambda}^C =  \eta_\text{p} \boldsymbol{\Psi}^\eta,
\end{equation}
where $C$ is the elastic modulus and $\eta_\text{p}$ is the polymeric viscosity.  The total deformation field $\mathbf{u}(\mathbf{r}) = \mathbf{u}^C(\mathbf{r}) + \mathbf{u}^\eta(\mathbf{r})$ is the sum of the elastic and viscous contributions, and it is related to the velocity field by $\mathbf{v}(\mathbf{r}) = \partial_t \mathbf{u}(\mathbf{r})$.  The symmetrized deformation and velocity gradients appearing in Equation \ref{eqlinve} are $\Lambda_{ij}^C = \frac{1}{2}\left(\partial_i u_j^C + \partial_ju_i^C \right)$ and $\Psi_{ij}^\eta = \frac{1}{2}\left(\partial_iv_j^\eta + \partial_j v_i^\eta \right)$.  We further have $\boldsymbol{\Psi}^C = \partial_t\boldsymbol{\Lambda}^C$ and $\boldsymbol{\Psi} = \boldsymbol{\Psi}^C + \boldsymbol{\Psi}^\eta$, allowing us to write
\begin{equation}
    \partial_t \sigma_{ij}^\text{p} = C \Psi_{ij} - \frac{C}{\eta_\text{p}}\sigma_{ij}^\text{p}. \label{eqMaxLine}
\end{equation}
This last equation is of the desired form for implementation, as it relates the total polymeric stress $\boldsymbol{\sigma}^\text{p}$ to the total strain rate tensor $\boldsymbol{\Psi}$, both of which are tracked in simulation.  

As a final step, we promote the partial derivative $\partial_t$ to an operator $\mathcal{D}_t$ which is materially objective, in the sense that rigid transformations of the coordinate system leave the dynamical equation unchanged \cite{phan2013understanding, bird1987dynamics, larson2013constitutive}.  Here we use the corotational (or Jaumann) derivative 
\begin{equation}
    \mathcal{D}_t X_{ij} = D_t X_{ij} + \Omega_{ik}X_{kj} - X_{ik}\Omega_{kj},
\end{equation}
where 
\begin{equation}
   \Omega_{ij} = \frac{1}{2}\left(\partial_i v_j - \partial_j v_i \right)
\end{equation}
represents the local vorticity.  Using the corotational derivative in Equation \ref{eqMaxLine} yields the Johnson-Segalman model \cite{oldroyd1950formulation, olmsted2000johnson}.  An important feature of spatially extended viscoelastic materials is the diffusion of the stress $\boldsymbol{\sigma}^\text{p}$ throughout the system volume \cite{mohammadigoushki2016flow,malek2018thermodynamics}.  This effect is captured in the diffusive Johnson-Segalman model\cite{olmsted2000johnson} \begin{equation}
    \mathcal{D}_t \sigma_{ij}^\text{p} =  C \Psi_{ij} - \frac{C}{\eta_\text{p}}\sigma_{ij}^\text{p} + D_\text{p} \partial_{kk} \sigma_{ij}^\text{p}, \label{eqMaxDiff}
\end{equation}
where $\partial_{kk}$ represents the Laplacian operator and $D_\text{p}$ is a diffusion constant describing the spreading of stress sustained by the viscoelastic medium.  We note that this diffusive term can improve model stability in some contexts by reducing gradients in the polymeric stress \cite{dzanic2022hybrid,liu2021viscoelastic,gupta2019effect}.  We show later that the viscoelastic return of a dragged droplet depends non-monotonically on $D_\text{p}$.

\begin{figure*}
\begin{center}
\includegraphics[width=\textwidth]{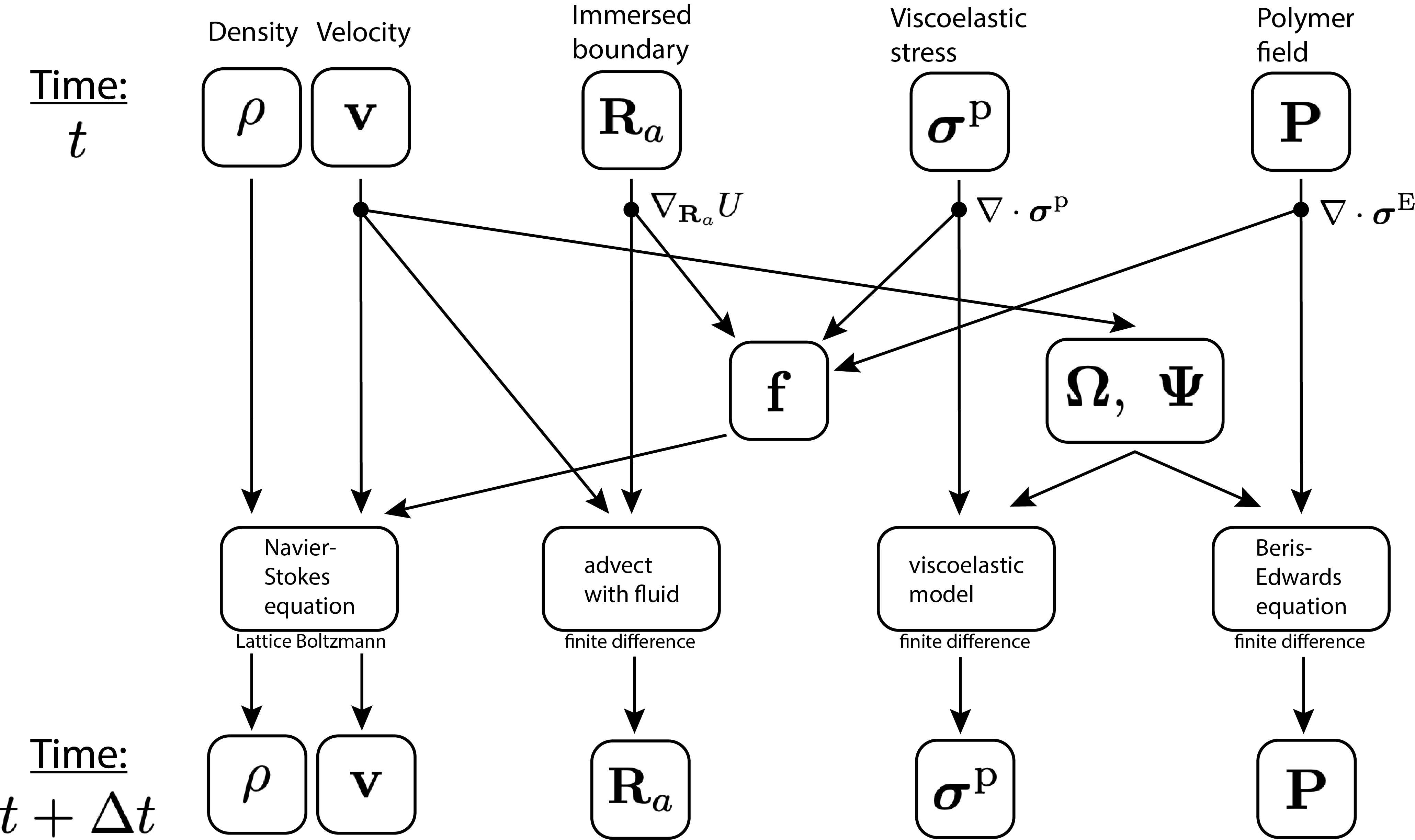}
\caption{Simulation flowchart. We show the various fields tracked during simulation and how they influence each others' dynamics, as well as the numerical methods used to update the fields from time $t$ to $t + \Delta t$.  The fields $\rho$ and $\mathbf{v}$ are updated using the lattice Boltzmann algorithm, while the remaining fields $\mathbf{R}_a$, $\boldsymbol{\sigma}^\text{p}$, and $\mathbf{P}$ are updated using finite difference methods.  A force $\mathbf{f}$ which enters into the lattice Boltzmann step takes contributions from three fields, and the field $\mathbf{v}$ in turn influences the dynamics of those three fields both directly and through the tensors $\boldsymbol{\Omega}$ and $\boldsymbol{\Psi}$.  Not visualized here is how $\mathbf{P}$ also affects the update of $\boldsymbol{\sigma}^\text{p}$ through the stiffness tensor $\mathbf{C}(\mathbf{P})$.  Because all fields are updated using inputs from the previous timestep, the order of updates does not matter.}
\label{Flowchart}
\end{center}
\end{figure*}

To numerically integrate Equation \ref{eqMaxDiff}, we use a finite difference predictor-corrector scheme with the same timestep  $\Delta t$ as the LB algorithm.  The viscoelastic stress tensor is defined on the same fixed grid as the fluid velocity and density.  This scheme---in which additional physical fields are integrated via finite difference in tandem with the LB algorithm---is known as a hybrid Lattice Boltzmann scheme \cite{mezrhab2004hybrid, carenza2019lattice}.  The algorithm described here is schematically illustrated in Figure \ref{Flowchart}.  In the next section, we use a similar hybrid scheme to track the evolution of a polymeric orientation field $\mathbf{P}(\mathbf{r})$, allowing us to progress beyond the scalar viscoelasticity presented here into a new class of models in which the viscoelastic forces throughout the fluid depend on the local polymer orientation through an elasticity tensor $\mathbf{C}(\mathbf{P})$.

\subsection{Tensorial viscoelastic model}\label{sec:TVM}
Because the viscoelastic properties of a structured fluid can depend on the local configuration of solvated polymers, we extend the hydrodynamic description of the system to account for the polymer orientation dynamics.  Our model of the dynamics builds on previous work on the hydrodynamics of polar gels \cite{marchetti2013hydrodynamics, carenza2019lattice, tjhung2012spontaneous}.  A common focus of this literature is on active polymer gels, which contain additional contributions to the stress tensor arising from chemical energy consumption.  Although our theory could be extended to include these contributions, we omit them here in order to focus on the tensorial character of the local elasticity and how the stiffness tensor depends on the polymer orientation, which were not treated previously.

\subsubsection{Polymer field dynamics}
In this subsection we summarize the previously established hydrodynamic equations governing the evolution of the polymer orientation field $\mathbf{P}(\mathbf{r},t)$, primarily following Ref.\ \citenum{carenza2019lattice}.  We then introduce our model, which allows the elasticity tensor to depend on $\mathbf{P}$, and an efficient parameterization.  

Let $\hat{\mathbf{n}}$ represent a unit vector pointing along a polymer's local tangent.  At a coarse-grained level of description in two dimensions, a point $\mathbf{r}$ in space is characterized by a local distribution $g(\hat{\mathbf{n}};\mathbf{r})$ of $\hat{\mathbf{n}}$ over the unit circle.  This distribution has a first moment equal to the local polarization vector 
\begin{equation}
    P_i(\mathbf{r}) = \int_{0}^{2 \pi} \hat{n}_i g(\hat{\mathbf{n}};\mathbf{r}) d\theta,
\end{equation}
where $\hat{\mathbf{n}} = \left(\cos\theta, \sin\theta \right)$ is specified by a polar angle $\theta$.  In three dimensions, this integral would be taken over the unit sphere and the differential solid angle $d\theta$ would be replaced by the solid angle. In the hydrodynamic theory, only the local polarization vector $\mathbf{P}(\mathbf{r})$ is tracked, and the full distribution $g(\hat{\mathbf{n}};\mathbf{r})$ is not known.  The magnitude $P$ of the polarization vector varies from $0$ to $1$ as the local distribution $g(\hat{\mathbf{n}})$ goes from a uniform distribution on the circle to a $\delta$ function.  The time evolution of field $\mathbf{P}(\mathbf{r},t)$ is governed by the Beris-Edwards equation \cite{beris1994thermodynamics, carenza2019lattice}, which for this system takes the form
\begin{equation}
    D_t P_i = -\Omega_{ik}P_k + \xi \Psi_{ik}P_k - \Gamma h_i(\mathbf{P}). \label{eqBE}
\end{equation}
Here, $\boldsymbol{\Psi}$ and $\boldsymbol{\Omega}$ are the symmetric and antisymmetric parts of the velocity gradient tensor $\nabla \mathbf{v}$, $\xi$ is a scalar flow-alignment parameter, $\Gamma$ is a rotational-diffusion constant, and $\mathbf{h}(\mathbf{P}) = \delta F/\delta \mathbf{P}$ is the molecular field, which is derived from a model-specific free energy functional $F[\mathbf{P}]$.  In this work, we take $F[\mathbf{P}]$ to be \cite{carenza2019lattice,tjhung2012spontaneous}
\begin{equation}
    F[\mathbf{P}] = \int d \mathbf{r} \left\{ \frac{\alpha}{2} \lvert \mathbf{P}\rvert^2 +  \frac{\beta}{4} \lvert\mathbf{P}\rvert^4 + \frac{\kappa}{2} \left( \nabla \mathbf{P} \right)^2 \right\}, \label{eqFfunctional}
\end{equation}
giving rise to the molecular field 
\begin{equation}
    h_i(\mathbf{P}) = \left(\alpha  + \beta P^2 \right) P_i - \kappa \partial_{kk} P_{i}.
\end{equation}
The coefficients $\alpha$ and $\beta$ control the isotropic ($P=0$) to polar ($P>0$) transition, while $\kappa$ controls the energetic cost of deformations from the aligned phase.  The fluid flow field influences the dynamics of $\mathbf{P}$ via Equation \ref{eqBE}, and $\mathbf{P}$ influences the flow field via an additional contribution to the fluid stress called the Ericksen stress tensor $\boldsymbol{\sigma}^\text{E}$, which for this system is\cite{carenza2019lattice}
\begin{equation}
    \sigma_{ij}^\text{E} = -\frac{1}{2}\left( P_i h_j - h_i P_j\right) + \frac{\xi}{2}\left( P_i h_j + h_i P_j\right) - \partial_i P_k \partial_j P_k .
\end{equation}
The divergence of this stress tensor is a force density, $f_i^\text{E} = \partial_j \sigma_{ij}^\text{E}$, which enters on the right hand side of Equation \ref{eqNS} and is handled in simulation similarly to other force contributions described earlier.  In our hybrid LB implementation, the Beris-Edwards equation is numerically integrated in tandem with the LB algorithm using the predictor-corrector method.

\subsubsection{Elasticity tensor of the polymer field}
Instead of a scalar elasticity modulus $C$, we now assume that the elastic response to deformation is characterized by a rank-four stiffness tensor $\mathbf{C}$, defined by the relation 
\begin{equation}
    \sigma_{ij}^\text{p} =  C_{ijkl} \Lambda_{kl}^C, \label{eqSigmaConst}
\end{equation}
where $\boldsymbol{\sigma}^\text{p}$ is the elastic stress tensor experienced in response to a deformation gradient $\boldsymbol{\Lambda}^C$.  Incorporating this tensorial element into the Maxwell model, we generalize Equation \ref{eqMaxDiff} as
\begin{equation}
     \mathcal{D}_t \sigma_{ij}^\text{p} =  C_{ijkl}\Psi_{kl} - \frac{1}{\eta_\text{p}} C_{ijkl}\sigma_{kl}^\text{p} + D_\text{p} \partial_{kk} \sigma_{ij}^\text{p}. \label{eqStiffCons}
\end{equation}
In principle the viscous response of the polymers may also require a tensorial description $\eta_{\text{p},ijkl}$, causing the second term in Equation \ref{eqStiffCons} to depend on a separate tensor formed from the elasticity and viscosity tensors ($R_{ijkl}$ in Ref.\ \citenum{banerjee2021active}).  This tensor could be straightforwardly accommodated by our model, but we currently neglect it for simplicity. 

We next describe how to express the dependence of the stiffness tensor $\mathbf{C}(\mathbf{P})$ on the local polymer polarization vector $\mathbf{P}(\mathbf{r})$.  The method presented here is based partly on the work of Kwon and coworkers described in Ref.\ \citenum{kwon2008microstructurally}, which itself builds on the work of Mackintosh and coworkers described in Refs. \citenum{mackintosh1995elasticity, storm2005nonlinear, broedersz2014modeling}.  The stiffness tensor is expressed as an integral over the local filament orientation distribution:
\begin{equation}
    C_{ijkl} = (\rho_\text{p} -\rho_\text{ref})^a \int_0^{2\pi}g(\hat{\mathbf{n}}) K_{ijkl}\left(\hat{\mathbf{n}}\right)d\theta, \label{eqCTens}
\end{equation}
where the integral is over the unit circle, and $\mathbf{K}$ is a separate rank four tensor discussed below.  The term $(\rho_\text{p} - \rho_\text{ref})^a$, which depends on the polymer density $\rho_\text{p}$ and two parameters $\rho_\text{ref}$ and $a$, allows for non-affine deformations of the fluid \cite{kwon2008microstructurally}.  In the affine case, the expression for $\mathbf{C}$ reduces to the one originally derived by Mackintosh and coworkers:
\begin{equation}
    C_{ijkl} = \rho_\text{p} K_{||} \int_0^{2\pi}g(\hat{\mathbf{n}}) \hat{n}_i\hat{n}_j\hat{n}_k\hat{n}_l d\theta, \label{eqCTensaff}
\end{equation}
where $K_{||}$ is the extensional stiffness of the polymer.  Thus in the affine case, $\rho_\text{ref} = 0$, $a = 1$, and $K_{ijkl} = K_{||} \hat{n}_i\hat{n}_j\hat{n}_k\hat{n}_l$.  The physical basis for Equations \ref{eqCTens} and \ref{eqCTensaff} is the Irving-Kirkwood formula for the stress tensor, as detailed in Refs.\   \citenum{kwon2008microstructurally}, \citenum{storm2005nonlinear}, and \citenum{irving1950statistical}.  

The tensor $\mathbf{K}(\hat{\mathbf{n}})$, can be expressed as a rotation of the constant tensor $\mathbf{K}^\text{x} = \mathbf{K}(\hat{\mathbf{x}})$ representing the elastic response of cross-linked polymers oriented along the unit vector $\hat{\mathbf{x}}$ in the direction of the $x$-axis.  The non-zero elements of $\mathbf{K}^\text{x}$ reflect its material symmetry properties, which we assume here to be transversely isotropic \cite{kwon2008microstructurally}.  In this symmetry class, a deformation along $\hat{\mathbf{x}}$ can produce transverse forces in the $\hat{\mathbf{y}}$ direction through non-affine cross-linking of the polymers.  To gain insight into the meaning of the elements of $\mathbf{K}^\text{x}$, we consider a polymer field oriented purely along the $x$-axis, so that $\mathbf{C} = (\rho_\text{p} -\rho_\text{ref})^a\mathbf{K}^\text{x}$.  We proceed in the Mandel basis, in which rank four tensors are expressed as matrices in $\mathbb{R}^{3\times 3}$ and matrix multiplication is used to represent contraction over multiple indices \cite{slaughter2012linearized,mazdziarz2019comment}.  We denote $\mathbf{K}$ in its Mandel basis as $\widetilde{\mathbf{K}}$, and similarly for other tensors.  The constitutive relation for the aligned polymer field is expressed in this basis as (cf.\ Equation \ref{eqSigmaConst})
\begin{equation}
    \widetilde{\sigma}_{\alpha}^C = (\rho_\text{p} -\rho_\text{ref})^a \widetilde{K}_{\alpha\beta}^\text{x} \widetilde{\Lambda}^C_\beta,
\end{equation}
where we use Greek letters as indices in the Mandel basis.  Here, $\widetilde{\boldsymbol{\sigma}}^C = (\sigma_{11}, \sigma_{22}, \sqrt{2}\sigma_{12})^\intercal$, $\widetilde{\boldsymbol{\Lambda}}^C = (\Lambda^C_{11}, \Lambda^C_{22}, \sqrt{2}\Lambda^C_{12})^\intercal$, and under the assumption of transverse isotropy, the matrix $\widetilde{\mathbf{K}}^\text{x}$ is given as\cite{mazdziarz2019comment}
\begin{equation}
    \widetilde{\mathbf{K}}^\text{x} = \begin{pmatrix}
    K^\text{x}_{1111} & K^\text{x}_{1122} & 0 \\
    K^\text{x}_{1122} & K^\text{x}_{2222} & 0 \\
    0 & 0 & 2K^\text{x}_{1212}
    \end{pmatrix}. \label{eqKmatrix}
\end{equation}
In this representation one can interpret the elements of $\widetilde{\mathbf{K}}^\text{x}$ as determining the elastic stress in response to various types of deformations: $K^\text{x}_{1111}$ is the longitudinal normal stiffness (along $\hat{\mathbf{x}}$) and $K^\text{x}_{2222}$ is the transverse normal stiffness (along $\hat{\mathbf{y}}$).  $K^\text{x}_{1122}$ represents the reciprocal force generated in one direction in response to a deformation along the orthogonal direction, and $K^\text{x}_{1212}$ represents an elastic resistance to shearing deformations.

Now let $\mathbf{Q}(\hat{\mathbf{n}})$ be the orthogonal tensor that rotates $\hat{\mathbf{x}}$ to $\hat{\mathbf{n}}$, i.e., $\hat{n}_i = Q_{ij}\hat{x}_j$; an expression for $\mathbf{Q}(\hat{\mathbf{n}})$ can be obtained in terms of the polar coordinate $\theta$ of $\hat{\mathbf{n}}$.  With this, $\mathbf{K}(\hat{\mathbf{n}})$ can be written in the original basis as (suppressing the dependence on $\hat{\mathbf{n}}$)
\begin{equation}
	K_{ijkl} = Q_{iI}Q_{jJ}Q_{kK}Q_{lL}K^\text{x}_{IJKL} 
	\label{eqrotate}
\end{equation}
with summation over repeated capital indices implied.  The rotation tensor $\mathbf{Q}$ and its inverse $\mathbf{Q}^{-1}$ have Mandel basis representations $\widetilde{\mathbf{Q}}$ and $\widetilde{\mathbf{Q}}^{-1}$, for which explicit formulas are available \cite{bao2005analysis} (cf.\ Equation \ref{qmatrix}).  In this basis the rotation in Equation \ref{eqrotate} is equivalent to
\begin{equation}
	\widetilde{K}_{\alpha \beta}(\hat{\mathbf{n}}) = \widetilde{Q}_{\alpha \gamma}(\hat{\mathbf{n}}) \widetilde{K}_{\gamma\delta}^\text{x} \widetilde{Q}_{\delta \beta}^{-1}(\hat{\mathbf{n}}), \label{eqKMandel}
\end{equation}
providing an expression for $\widetilde{\mathbf{K}}(\hat{\mathbf{n}})$ in terms of the elastic constants in $\widetilde{\mathbf{K}}^\text{x}$ and the orientation of the polymers $\hat{\mathbf{n}}$.

\subsubsection{Parameterizing the distribution of polymer orientations}
\begin{figure}[bt]
\begin{center}
\includegraphics[width=\columnwidth]{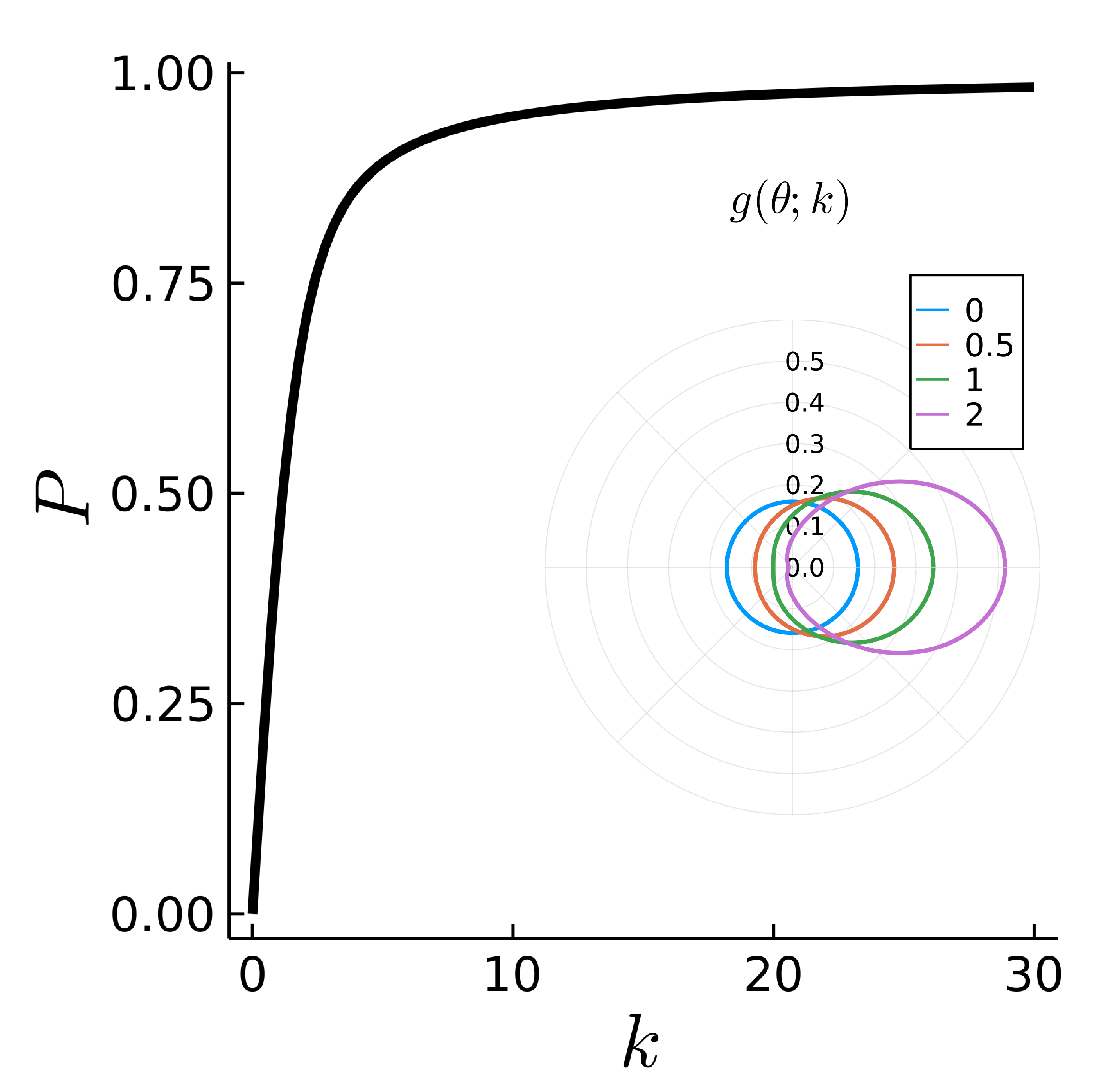}
\caption{Plot of the function $P$ in Equation \ref{eqPk}.  In the inset, a polar plot of the von-Mises distribution $g(\theta;k)$ is shown for four values of $k$ ranging from $0$ to $2$.  In the polar plot, the value of $g(\theta;k)$  is represented by the distance from the origin to the curve at the corresponding value of $\theta$.}
\label{StiffTens}
\end{center}
\end{figure}

We next consider the distribution of polymer orientations, which we now write as a function $g(\theta)$ of the polar angle $\theta$ of $\hat{\mathbf{n}}$.  Because we track the first moment of this distribution $\mathbf{P}$ in the hydrodynamic model, we wish to find an expression for $g(\theta)$ consistent with $\mathbf{P}$ to use in Equation \ref{eqCTens}.  We consider the magnitude $P = \lvert \mathbf{P} \rvert$ and orientation $\hat{\mathbf{P}} = \mathbf{P} / P$ separately.  We first consider the special case that $\mathbf{P}$ lies along the $x$-axis (i.e., $\hat{\mathbf{P}} = \hat{\mathbf{x}}$).  The polarization $P$ is then obtained from $g(\theta)$ as the mean of the projection along $\hat{\mathbf{x}}$:
\begin{equation}
    P = \int_0^{2 \pi} g(\theta) \cos(\theta) d\theta. \label{eqPtheta}
\end{equation}
Viewing Equation \ref{eqPtheta} as a constraint on the possible distributions $g(\theta)$, together with the constraint that it is normalized, we now make the physically motivated choice that $g(\theta)$ should maximize the entropy $-\int_0^{2\pi}g(\theta)\ln g(\theta)d\theta$.  It can be shown that the unique function with this property is the von-Mises distribution\cite{watson1982distributions}:
\begin{equation}
    g(\theta;k) = \frac{e^{k \cos\theta}}{2 \pi I_0(k)},
\end{equation}
where $I_0(k)$ is the modified Bessel function of order zero and $k >0 $ is a parameter that determines the anisotropy of the distribution.  This function is illustrated in the inset of Figure \ref{StiffTens} for different values of $k$.  To account for the general case when $\hat{\mathbf{P}} \neq \hat{\mathbf{x}}$, we simply find the polar angle $\theta_P$ of $\hat{\mathbf{P}}$ and write the density as $g(\theta - \theta_P;k)$.

In addition to its physical motivation, there are two key advantages of this choice for $g(\theta)$.  The first is that it allows the integral over $\theta$ in Equation \ref{eqCTens} to be performed analytically.  The second is that it allows for a one-to-one mapping between $P$ and $k$, which follows from the relation 
\begin{equation}
    P(k) = \int_0^{2 \pi} g(\theta;k) \cos(\theta) d\theta = \frac{I_1(k)}{I_0(k)}. 
    \label{eqPk}
\end{equation}
Although this relation (plotted in Figure \ref{StiffTens}) cannot be inverted analytically, $k$ can be readily determined from $P$ numerically.  We note that there is no value of $k$ for which $P = 1$, but in practice either the allowed range of $P$ can be restricted to the interval $[0, 1)$, or one can take the limit $k\rightarrow \infty$.  The equilibrium value of $P$ is determined by the parameters $\alpha$ and $\beta$ in Equation \ref{eqFfunctional}.  In this paper, we ensure through our choices of these parameters that the equilibrium value is less than 1, and we account for the possibility of advection causing $P > 1$ by setting a maximal value $P = 0.99$ when determining $k$ from Equation \ref{eqPk}.

We provide the full expression for $\widetilde{\mathbf{C}}$ obtained using this method in Appendix \ref{elasticityTensor}.  In the isotropic case, when $k = 0$, the expression for $\widetilde{\mathbf{C}}$ reduces to 
\begin{equation}
    \widetilde{\mathbf{C}} = \begin{pmatrix}
    \widetilde{C}_{11} & \widetilde{C}_{12} & 0 \\
    \widetilde{C}_{12} & \widetilde{C}_{11} & 0 \\
    0 & 0 & \widetilde{C}_{11} - \widetilde{C}_{12}
    \end{pmatrix},
\end{equation}
where 
\begin{align}
    \widetilde{C}_{11} =& \frac{1}{8} (3 K^\text{x}_{1111} + 2 K^\text{x}_{1122} + 4 K^\text{x}_{1212} + 3 K^\text{x}_{2222}) \\
    \widetilde{C}_{12} =& \frac{1}{8} (K^\text{x}_{1111} + 6 K^\text{x}_{1122} - 4 K^\text{x}_{1212} + K^\text{x}_{2222}).
\end{align}
This is the expected form for an isotropic elasticity tensor in the Mandel basis, as described in Ref. \citenum{mazdziarz2019comment}. By setting $\widetilde{C}_{12}$ to zero through particular choices of the elements of $\mathbf{K}^\text{x}$, a single parameter $\widetilde{C}_{11}$ characterizes the elastic response, and we recover the case of scalar elasticity described in Section \ref{sec:scalar}.

In summary, the local polarization vector $\mathbf{P}(\mathbf{r}) = P \hat{\mathbf{P}}$ determines the stiffness tensor $\widetilde{\mathbf{C}}(\mathbf{P})$ through the orientation parameter $\theta_P$, which depends on $\hat{\mathbf{P}}$, and the anisotropy parameter $k$, which depends on $P$.  These together specify the parameterized distribution $g(\theta - \theta_P;k)$, which enters in the integral in Equation \ref{eqCTens}.  Given the expression for $\widetilde{\mathbf{K}}(\hat{\mathbf{n}})$ in Equation \ref{eqKMandel}, the integral can be computed analytically.  The result is then used in Equation \ref{eqStiffCons}, which is numerically integrated using a finite difference method as described in Sections \ref{sec:scalar} and \ref{sec:TVM}.  

\section{Results}
Here we describe several test cases demonstrating the novel physical features which may be resolved using the simulation methods outlined above.  We simulate an optical trap experiment, in which an elastic droplet is dragged by a harmonic trap through a viscoelastic fluid and then released.  Throughout this section, the main metric we use to evaluate the behavior is the return, which is defined as 
\begin{equation}
    \text{Return} = 1 - d_\text{rec} / d_\text{pull}, \label{eqRec}
\end{equation}
where $d_\text{rec}$ is the distance from the lipid droplet's initial center of mass position to its position at the end of the simulation, and $d_\text{pull}$ is the length of the harmonic trap pulling protocol.  If the droplet fully returned to its starting position, $d_\text{rec} = 0$ and $\text{Return=1}$.  The parameterization of the simulations and other implementation details are described in Appendix \ref{Param}.

\begin{figure}[!ht]
\begin{center}
\includegraphics[width=0.9\columnwidth]{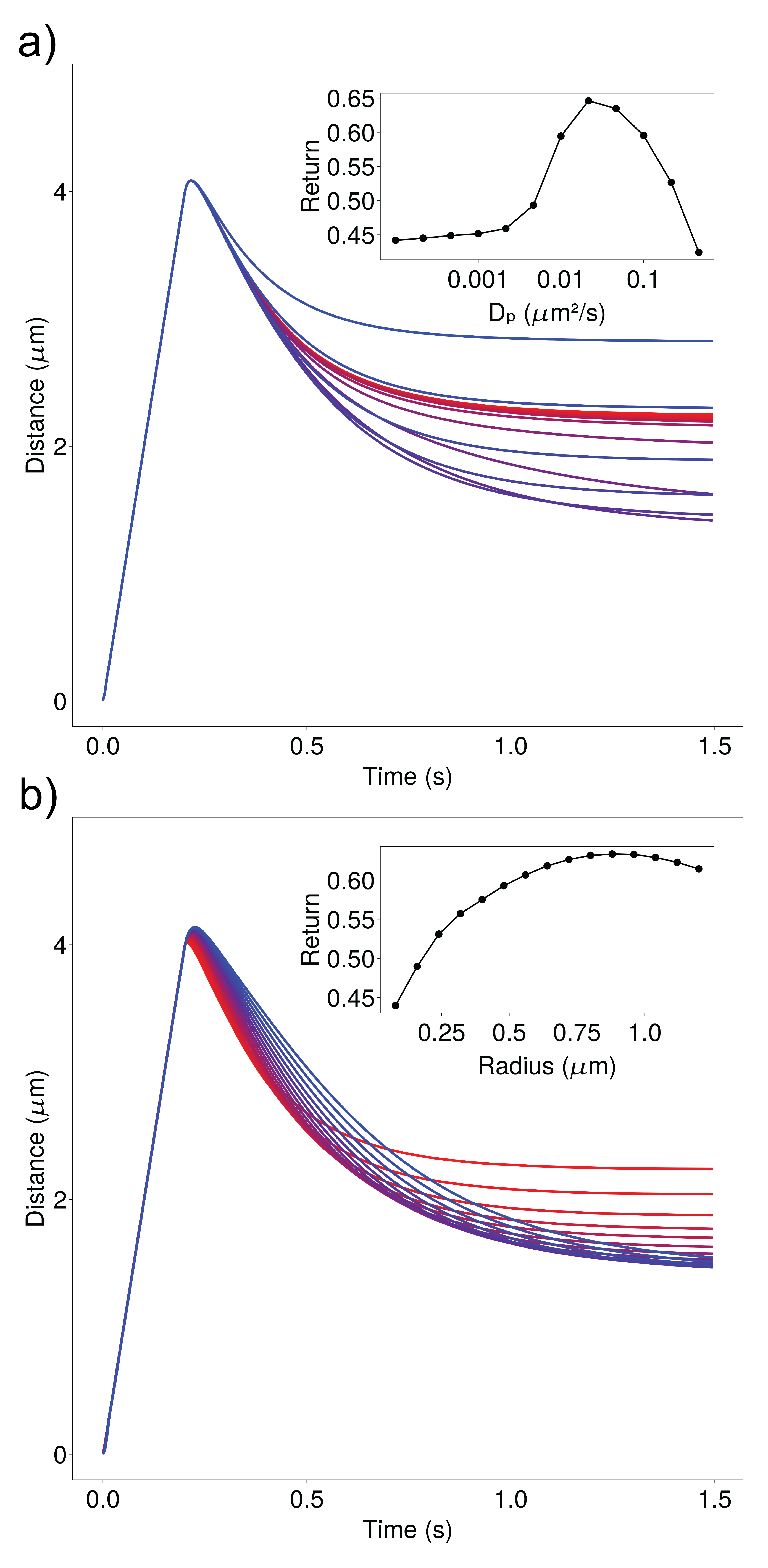}
\caption{Distance trajectory for several values of (a) the stress diffusion $D_\text{p}$ and (b) the droplet radius $R$. Insets: the value of the final return defined in Equation \ref{eqRec}.  The color ranges from red to blue in each plot as the parameters increase over the values at which points are plotted in the corresponding insets.  }
\label{ScalarResults}
\end{center}
\end{figure}

\subsection{Scalar viscoelasticity results}

\begin{figure*}[hbt]
\begin{center}
\includegraphics[width=\textwidth]{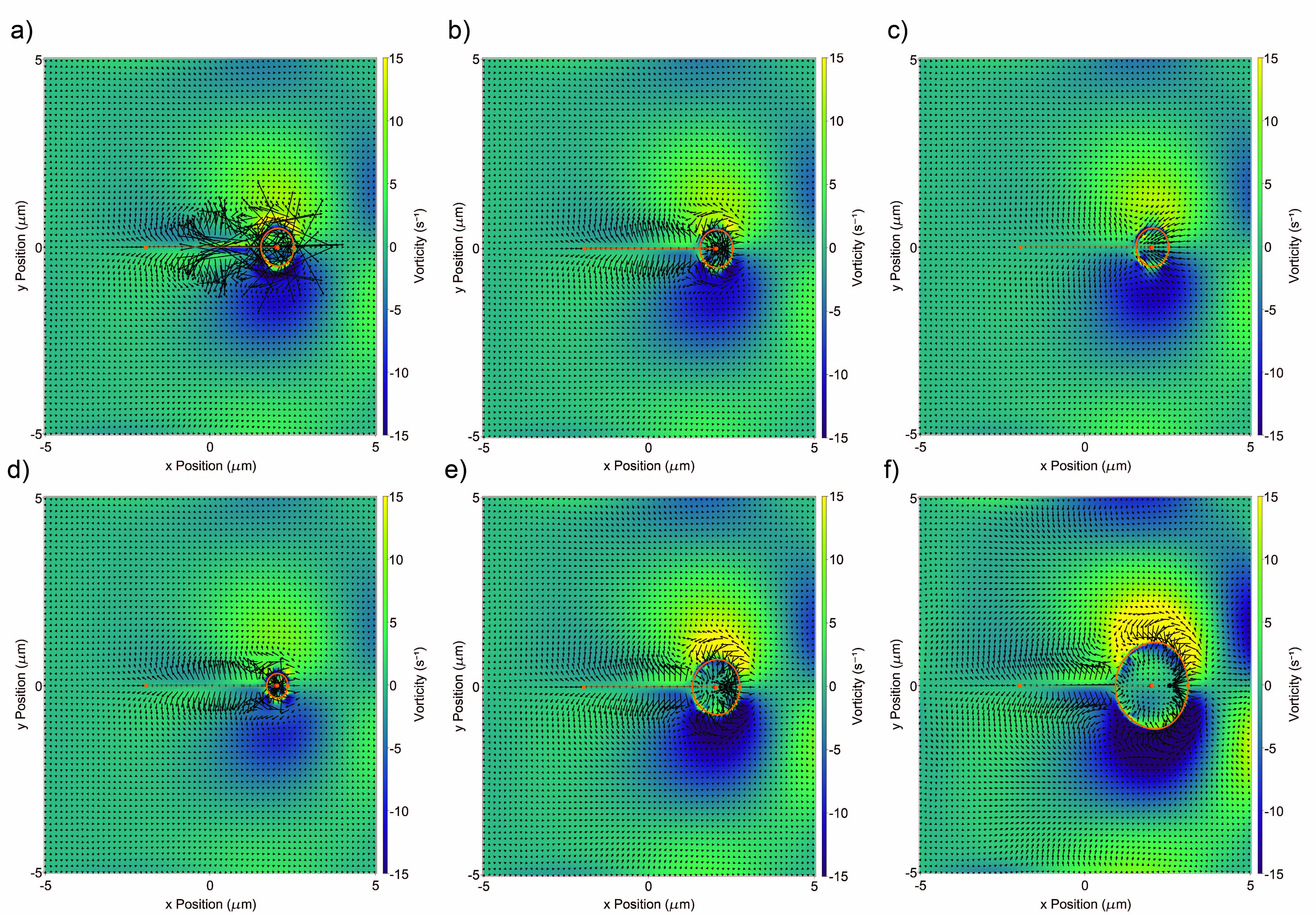}
\caption{Example responses to a driven body.  Simulations with (a) $D_\text{p} = 0.001 \ \mu\text{m}^2/\text{s}$ and $R = 0.5 \ \mu \text{m}$, (b) $D_\text{p}= 0.1 \ \mu\text{m}^2/\text{s}$ and $R = 0.5 \ \mu,  \text{m}$, (c) $D_\text{p}= 1 \ \mu\text{m}^2/\text{s}$ and $R = 0.5 \ \mu  \text{m}$, (d) $D_\text{p} =0.1 \ \mu\text{m}^2/\text{s}$ and $R = 0.32 \ \mu\text{m}$, (e) $D_\text{p} =0.1 \ \mu\text{m}^2/\text{s}$ and $R = 0.72 \ \mu\text{m}$,  and (f) $D_\text{p} = 0.1 \ \mu\text{m}^2/\text{s}$ and $R = 1.12 \ \mu\text{m}$.  In these images, the color represents the local vorticity of the fluid flow field $\partial_x v_y - \partial_y v_x$, and the arrows, shown at $1/16^\text{th}$ of grid points, are proportional to the local vector $\nabla\cdot \boldsymbol{\sigma}^\text{p}$, which represents the viscoelastic restoring force at each point.  The thick orange curve represents the boundary of the droplet, defined by the points $\{ \mathbf{R}_a \}_{a=1}^{N_{IB}}$ (cf.\ Figure \ref{Droplet}), and the three orange dots represent the initial center of mass, the current center of mass, and the position of a fixed point on the boundary.  The thin orange line traces the trajectory of the center of mass during the pulling protocol.  The red dot (which overlaps with the current center of mass) represents the location of the harmonic trap. All snapshots correspond to the time $t = 0.2 \ \text{s}$.}
\label{ScalarResultsSnapshots}
\end{center}
\end{figure*}

\begin{figure}[hbt]
\begin{center}
\includegraphics[width=0.9\columnwidth]{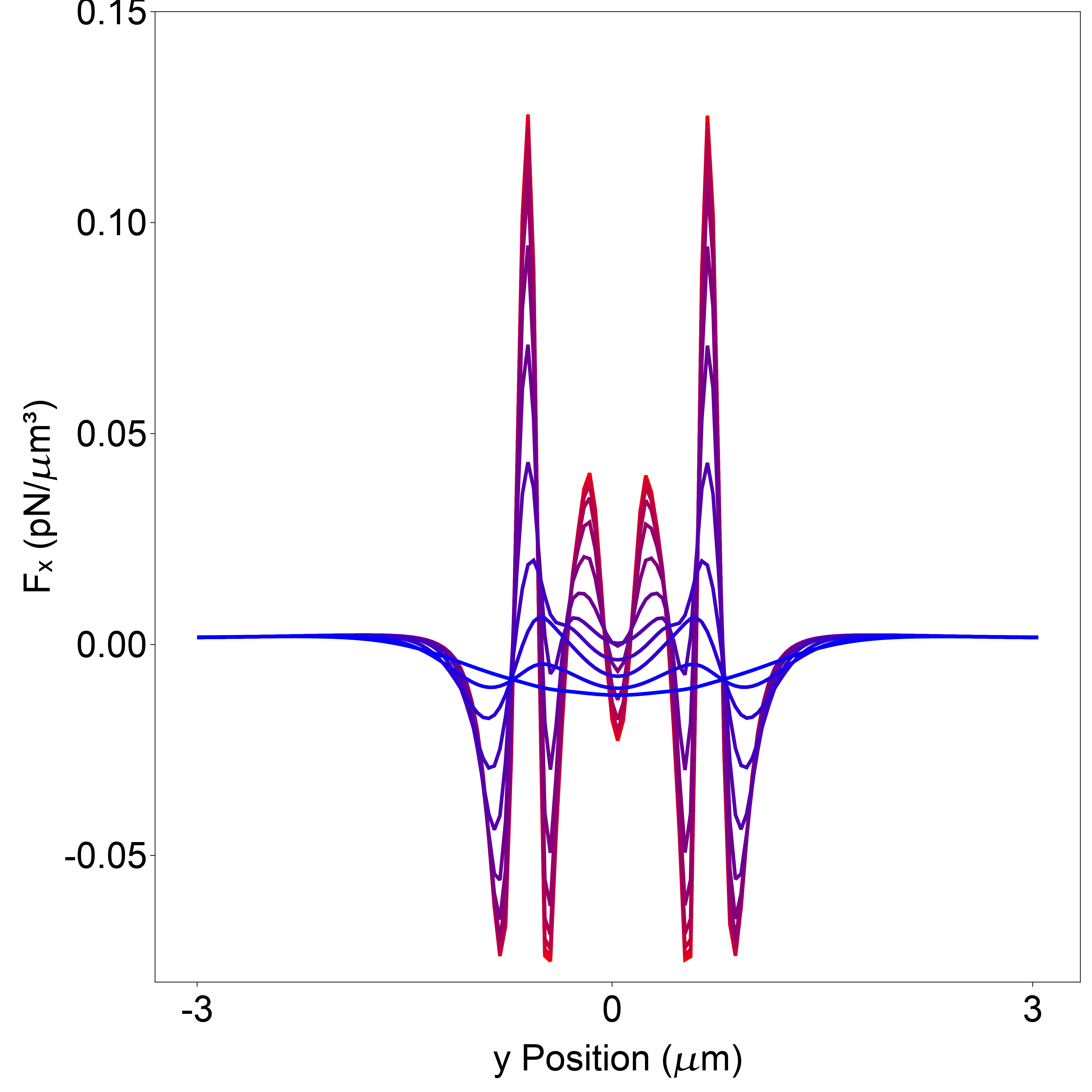}
\caption{Effect of stress diffusion on the force distribution.  For the values of $D_\text{p}$ used in Figure \ref{ScalarResults}a, the $x$-component of $\nabla \cdot\boldsymbol{\sigma}^\text{p}$ is visualized at $t = 0.2 \ \text{s}$ on the vertical line at $x = 2 \ \mu\text{m}$ passing through the droplet's center of mass.  The colors range from red to blue as $D_\text{p}$ increases, showing how the force distribution becomes more spatially inhomogeneous for small $D_\text{p}$.  For clarity only part of the full domain $[-5\ \mu\text{m}, \ 5\ \mu\text{m}]$ has been visualized.}
\label{DsFx}
\end{center}
\end{figure}

Several parameters entering the simulation setup are shown here to have significant effects on the return.  In Figure \ref{ScalarResults} we show how the return depends on two such parameters: the droplet radius $R$ and the stress diffusion constant $D_\text{p}$. The range of $D_\text{p}$ is consistent in order of magnitude with the measurements made in Ref.\ \citenum{mohammadigoushki2016flow} on the viscoelastic stress diffusion in a worm-like micellar solution of CTAB-NaNO$_3$. We observe a non-monotonic dependence of the return on each of these.  In the case of $D_\text{p}$ (Figure \ref{ScalarResults}a), this non-monotonic dependence can be understood as follows.  At very low values of $D_\text{p}$, the viscoelastic restoring force $\nabla\cdot \boldsymbol{\sigma}^\text{p}$ has large gradients and is spatially inhomogeneous.  In particular, significant components of the restoring force do not align with the the direction of return, which diminishes the return. The vector field corresponding to this case is shown in Figure \ref{ScalarResultsSnapshots}a.  On the other hand, when $D_\text{p}$ is large (Figure \ref{ScalarResultsSnapshots}b), the stress diffuses quickly, which diminishes the magnitude of $\nabla\cdot \boldsymbol{\sigma}^\text{p}$ and, in turn, the  return.  See Figure \ref{DsFx} for an illustration of these forces as $D_\text{p}$ is varied.  These competing effects cause the optimal return to occur at intermediate values of $D_\text{p}$.  The non-monotonic dependence on the radius (Figure \ref{ScalarResults}b) is a result of a trade-off between higher drag on the droplet and larger elastic displacement of the droplet as the radius increases (Figure \ref{ScalarResultsSnapshots}c-d).  See Supplementary Videos for movies of these simulations.

\begin{figure}
\begin{center}
\includegraphics[width= \columnwidth]{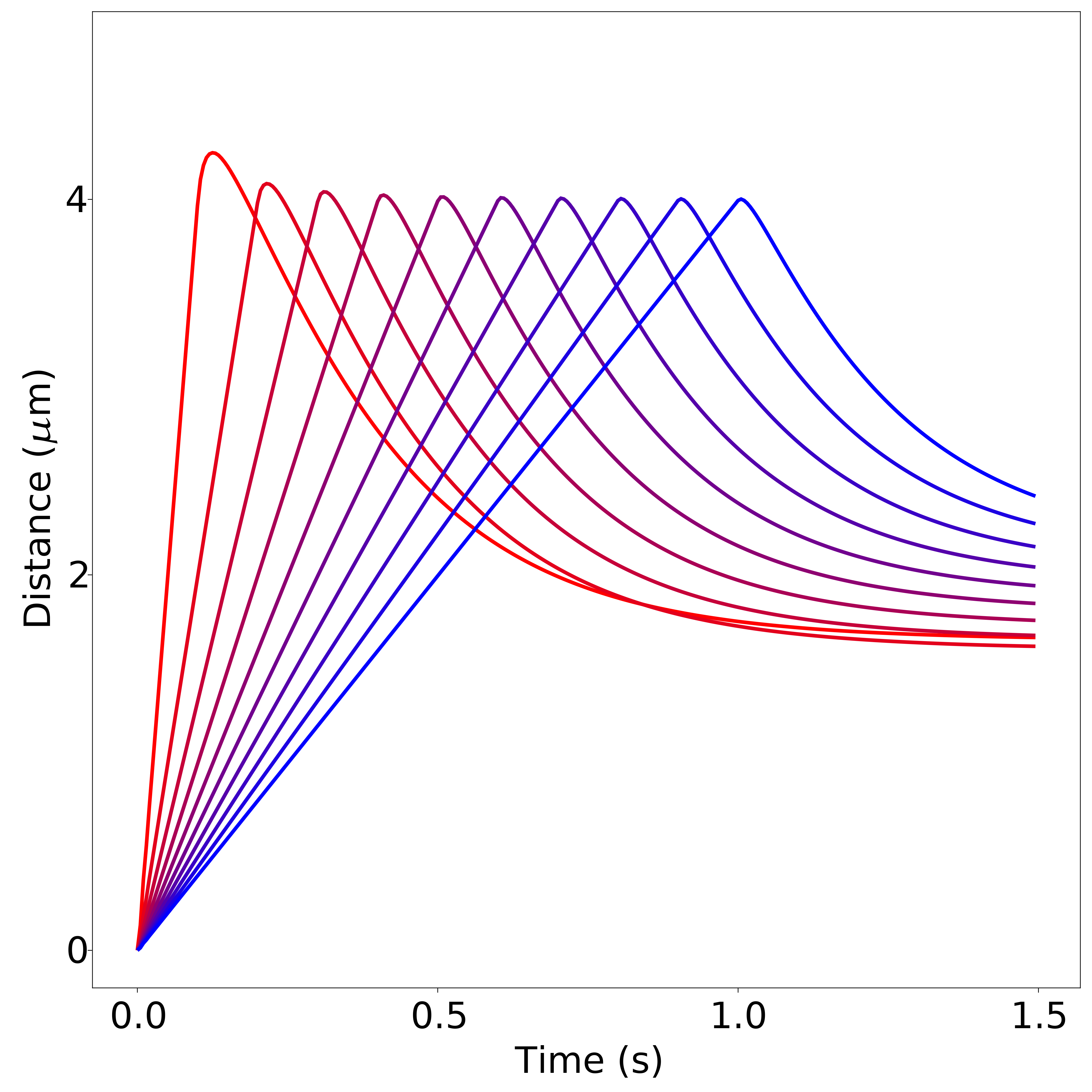}
\caption{Distance trajectories for several pulling speeds.  The time $T_\text{pull}$ over which the droplet is pulled is evident from point at which each trajectory obtains its maximum (except for the smallest values of $T_\text{pull}$ which exhibit some overshoot due to convective fluid motion).  }
\label{SpeedRecoil}
\end{center}
\end{figure}

In Figure \ref{SpeedRecoil}, we show trajectories of the droplet distance as we vary the time $T_\text{pull}$ over which the droplet is pulled the distance $d_\text{pull} = 4 \ \mu$m.  We see that as the droplet is pulled more quickly, it returns closer to its original position.  This can be attributed to the reduced amount of time over which the viscous dissipation acts to diminish the viscoelastic restoring force.

\subsection{Tensorial viscoelasticity results}

\begin{figure}
\begin{center}
\includegraphics[width= \columnwidth]{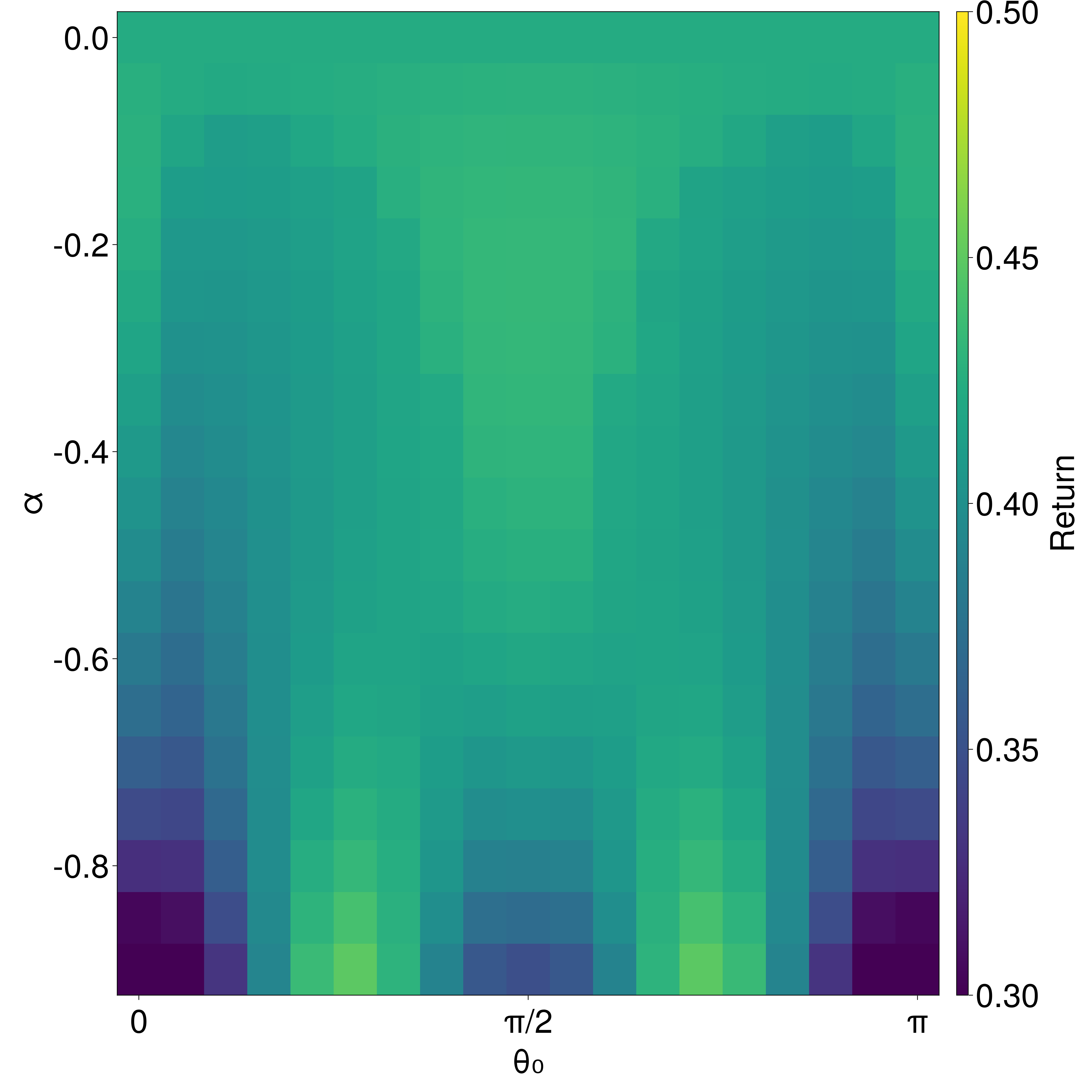}
\caption{The value of the return, defined in Equation \ref{eqRec}, as a function of the parameters $\alpha$, which controls the typical polarization $P$ through $P = \sqrt{-\alpha/\beta}$, and $\theta_0$, the initial angle between the aligned polymer field an the $+x$ direction.}
\label{AlphaTheta}
\end{center}
\end{figure}

We now illustrate how using a tensorial description of polymer elasticity can produce anisotropy and spatial asymmetry in the viscoelastic restoring forces.  The stiffness tensor $\mathbf{C}(\mathbf{P})$ has entries depending on the local polymer polarization vector $\mathbf{P}(\mathbf{r})$, such that both the magnitude of the polarization $P$ and the relative angle between the elastic deformation and the polymer orientation $\hat{\mathbf{P}}$ contribute to the viscoelastic response.  To illustrate this, we fix the direction of pulling along the $x$-axis and vary both the equilibrium polarization $P$ and the initial angle $\theta_0$ which the aligned polymer field makes with this direction.  The equilibrium polarization can be found from the free energy (cf.\ Equation \ref{eqFfunctional}) as $\sqrt{-\alpha / \beta}$, and we fix $\beta = 1$ and vary $\alpha$ from $0$ to $-0.9$, causing $P$ to vary from $0$ to approximately $0.95$.  The return as a function of $\alpha$ and $\theta_0$ is shown in Figure \ref{AlphaTheta}.  We see that, as expected, when $\alpha = P = 0$ the system is isotropic and there is no dependence of the return on $\theta_0$.  As $P$ grows, strong dependencies of the return on $\theta_0$ emerge, indicating the onset of anisotropic viscoelastic response.  We observe that this anisotropic response, as measured by the return parameter, is periodic with period $\pi$ and has a reflection symmetry about $\pi/2$.

\begin{figure*}
\begin{center}
\includegraphics[width= \textwidth]{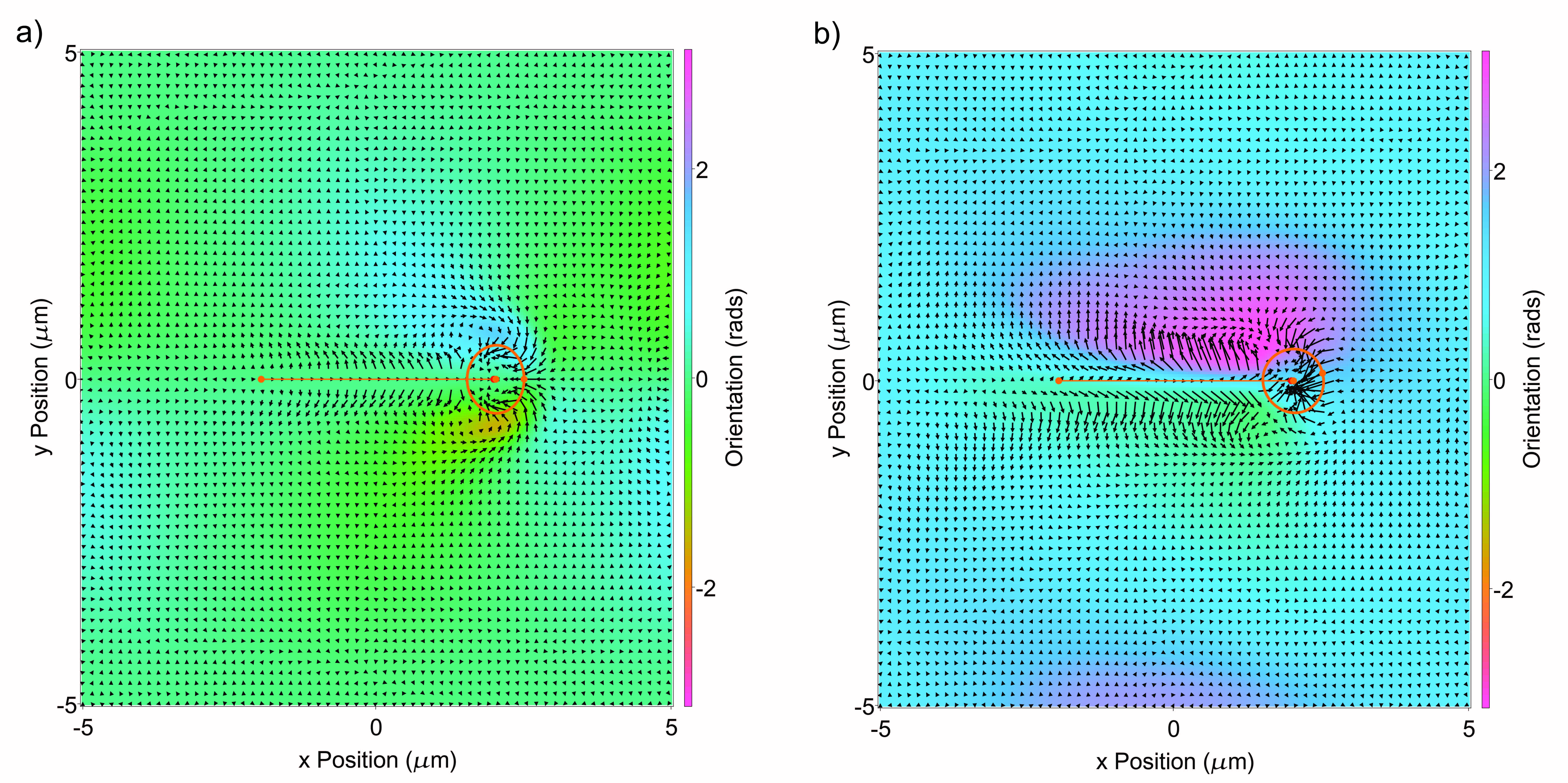}
\caption{The orientation field $\mathbf{P}(\mathbf{r})$ at $t = 0.2 \ \text{s}$ with initial orientation angle (a) $\theta_0 = 0$ and (b) $\pi/3$; $\alpha = -0.9$.  The color indicates the local angle which $\mathbf{P}(\mathbf{r})$ makes with the $x$-axis, and the arrows are proportional to $\nabla\cdot \boldsymbol{\sigma}^\text{p}$.  The remaining details of the visualization are the same as in Figure \ref{ScalarResultsSnapshots}.  Note that the tracked point represented by the orange dot on the droplet's boundary has rotated off of the line $y = 0 \ \mu\text{m}$ in panel b.  }
\label{OrientationFields}
\end{center}
\end{figure*}

The fluid flow from the dragged droplet distorts the initially aligned polymer field via the coupling of $\mathbf{P}(\mathbf{r})$ to the velocity gradient tensor (cf.\ Equation \ref{eqBE}).  This is visualized in Figure \ref{OrientationFields} for two initial orientations $\theta_0 = 0$ and $\pi/3$.  For $\theta_0 = 0$, the pulling direction and initial polymer orientation are parallel, and the restoring forces $\nabla\cdot \boldsymbol{\sigma}^\text{p}$ are symmetric about the pulling direction.  However, this symmetry is broken when $\theta_0 =\pi/3$, and as a result the restoring forces push asymmetrically on the droplet.  We observe that this asymmetry causes the droplet to deflect off the axis along which it was pulled, as well as to rotate slightly (see Supplementary Videos).

\begin{figure*}
\begin{center}
\includegraphics[width= \textwidth]{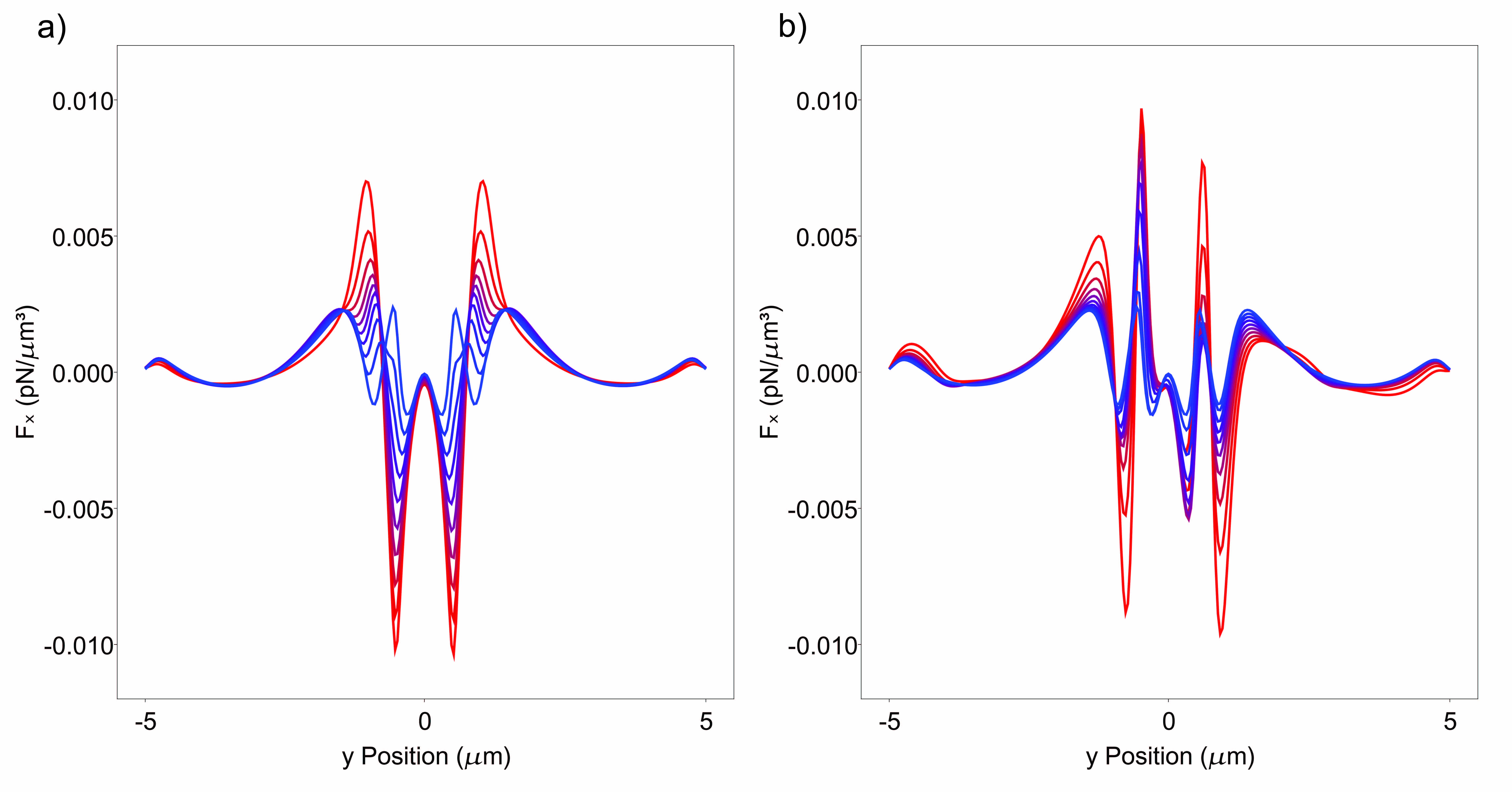}
\caption{The $x$-component of the force density $\nabla \cdot \boldsymbol{\sigma}^\text{p}$ on the line $x = 2 \ \mu\text{m}$ for (a) $\theta_0 = 0$ and (b) $\pi/3$ as $\alpha$ is ranges from $0$ (blue) to $-0.9$ (red).  These plots correspond to the same simulation and time point as shown in Figure \ref{OrientationFields}.  Note that asymmetries about $y = 0 \ \mu\text{m}$ develop for $\theta_0 = \pi/3$ when $\alpha <0$.  }
\label{FxComposite}
\end{center}
\end{figure*}

We highlight that in the tensorial model, spatial asymmetries of the restoring force which are not captured in the scalar model can now be resolved.  This is further illustrated in Figure \ref{FxComposite}, where we show the $x$-component of $\nabla \cdot \boldsymbol{\sigma}^\text{p}$ for a range of values of $\theta_0$ and $\alpha$.  For $\theta_0 = 0$, the restoring force is symmetric about $y = 0 \ \mu\text{m}$ for all values of $\alpha$, while for $\theta_0 = \pi / 3$ asymmetries develop as $\alpha$ increases.  These unbalanced restoring forces on the top and bottom of the droplet lead to its deflection and rotation, as noted above.

\section{Discussion}
Our goal in this work was to develop a method for treating fluid structure in hydrodynamic simulations of viscoelastic media.  In particular, we introduced a tensor representation of the elastic modulus $\mathbf{C}(\mathbf{P})$ that depends on the local polarization vector $\mathbf{P}(\mathbf{r})$.  In the test cases that we presented, this was shown to cause the viscoelastic return of a dragged lipid droplet to become strongly anisotropic.  In addition, it was shown that, even in the case that the elastic modulus is a scalar, the viscoelastic return depends non-monotonically on the droplet's size and the diffusion rate of the polymeric stress.   We note that, while scalar viscoelasticity has been treated in previous lattice Boltzmann works, we are not aware of any studies on the viscoelastic return accompanying a pulled droplet as reported here.  Further, the tensorial model of viscoelasticty has not been treated in previous works.  These applications demonstrate novel physical features which the modeling approaches developed in this work can resolve.

A number of extensions to the work presented here are possible.  First, generalizing these methods to three dimensions rather than two should be straightforward.  One difficulty which can arise is analytically evaluating the integral of the parameterized distribution $g(\hat{\mathbf{n}};k)$ over the unit sphere rather than the unit circle (cf.\ Equation \ref{eqCTens}).  The choice of $g(\hat{\mathbf{n}})$ here as a von-Mises distribution facilitated computing the stiffness tensor analytically in two dimensions, but computing the stiffness tensor in three dimensions may require resorting to numerical methods.  Second, the expression for the elastic stress tensor can be written in a more general setting as a double integral over a two-body distribution function of polymer orientations, rather than as a single integral over a one-body distribution $g(\hat{\mathbf{n}})$ as in Ref.\ \citenum{liao2019mechanism}.  This approach can allow incorporating domain knowledge of the polymeric system being simulated but will likely complicate the implementation.  Third, it has been argued that in addition to introducing a new timescale into the equation of motion for the stress tensor $\boldsymbol{\sigma}^\text{p}$, viscoelasticity also introduces a timescale into the Beris-Edwards equation for the dynamics of the polymer field \cite{marchetti2013hydrodynamics, joanny2009active}.  Incorporating this effect in the simulations presented here would be straightforward.  

Finally, we neglected here the possibility of active contractility of the viscoelastic fluid, which would add an additional contribution to the total stress tensor in the Navier-Stokes equation; including this feature should also be straightforward \cite{carenza2019lattice}.  In future work we aim to study the interplay between active stresses and the tensorial viscoelastic restoring forces which are the subject of this paper.  It would be interesting to allow the active stresses and elastic stiffness to depend on the local concentration of bound molecular motors in a force-dependent manner, capturing the catch bond-like dynamics of motors such as myosin II \cite{thomas2008biophysics}.  Because feedback loops between the chemistry and mechanics can produce complex energy landscapes, this should lead to a rich dynamical phase diagram with regions in which precise memories of the material's force history can be encoded.  

\section{Acknowledgments}

We wish to thank Fred Chang for helpful discussions that motivated the simulations presented.  This work was supported by the National Science Foundation through awards DMR-1848306, EF-1935260, and DMR-2011854. CF acknowledges support from the University of Chicago through a Chicago Center for Theoretical Chemistry Fellowship.  The authors acknowledge the University of Chicago’s Research Computing Center for computing resources.

\appendix

\section{Details of the immersed boundary method}\label{IBLBApp}
\begin{figure}
\begin{center}
\includegraphics[width= 0.9\columnwidth]{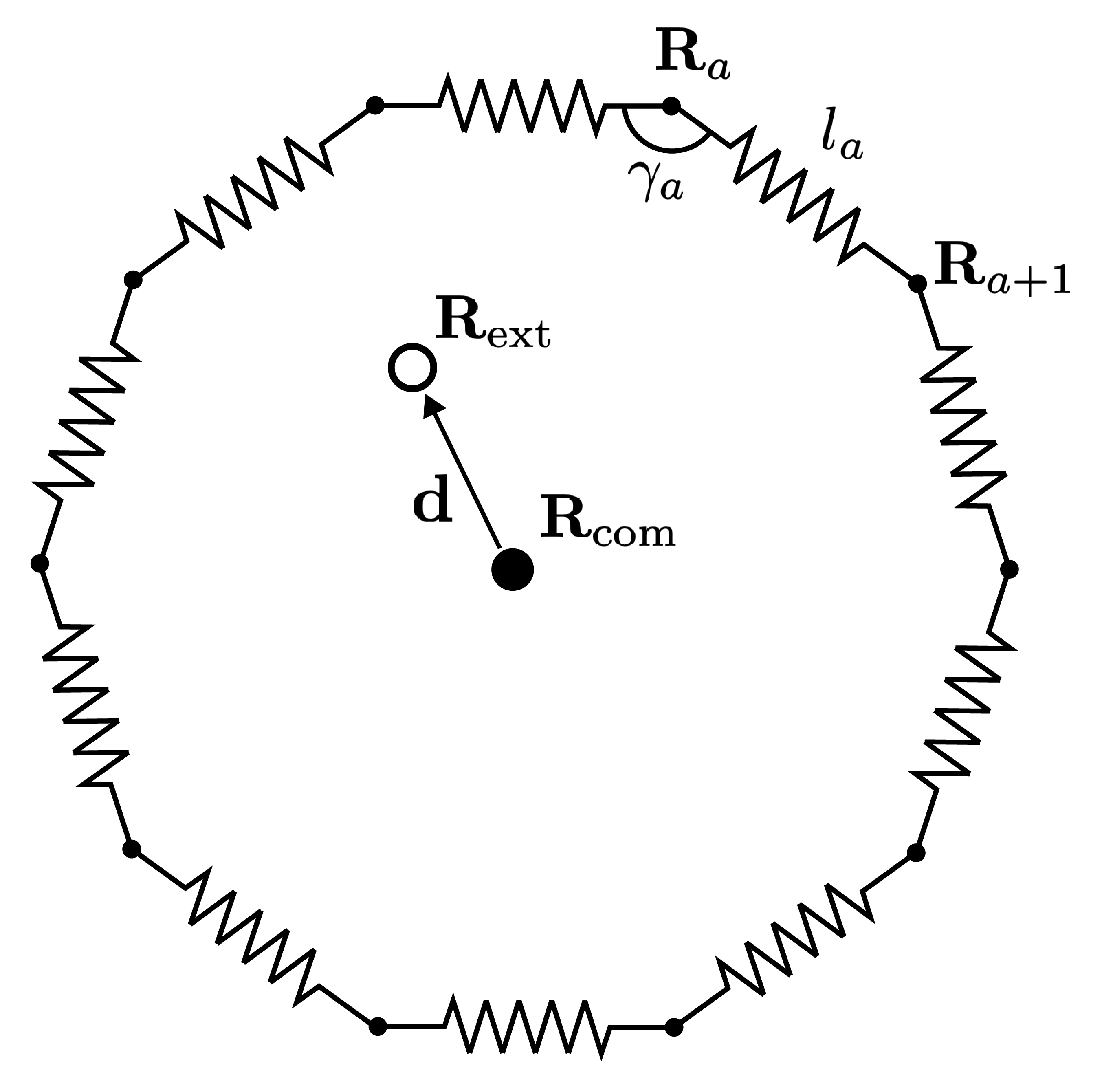}
\caption{The $N_{IB}$ points $\mathbf{R}_a$ that make up the immersed boundary are illustrated for $N_{IB} = 10$, showing the quantities $l_a$, $\gamma_a$, $\mathbf{R}_\text{com}$, $\mathbf{R}_{\text{ext}}$, and the trap displacement vector $\mathbf{d} = \mathbf{R}_{\text{ext}} - \mathbf{R}_\text{com}$ which enter into the constitutive equation for the boundary.}
\label{Droplet}
\end{center}
\end{figure}

The IB method was introduced by Peskin to model the interaction of fluids with an elastic boundary \cite{peskin1972flow, peskin2002immersed}.  The boundary is represented by a set of $N_{IB}$ points $\mathbf{R}_a$ which are not confined to the regular grid of the LB domain but can take any position within the simulation volume.  We use $a$ to index the $a^\text{th}$ point in the IB.  In the applications that we present, the points $\mathbf{R}_a$ represent the one-dimensional boundary of a two-dimensional droplet.  The elastic properties of the droplet are modeled through a constitutive equation describing the droplet's energy as a function of its configuration
\begin{equation}
    U\left(\{\mathbf{R}_a \}_{a=1}^{N_{IB}} \right) = \sum_{a=1}^{N_{IB}} \left( u_{a}^\text{spring} + u_a^\text{angle} \right) +  u^\text{trap}. \label{eqConst}
\end{equation}  The function $u_{a}^\text{spring}$ is a  harmonic potential that restrains the separation $l_a= \lvert  \mathbf{R}_{a+1} - \mathbf{R}_a \rvert$ between neighboring points to be close to an equilibrium value $l_\text{eq}$:
\begin{equation}
    u_{a}^\text{spring} = \frac{k_\text{spring}}{2}\left(l_a - l_\text{eq}\right)^2.
\end{equation}
The function $u_a^\text{angle}$ penalizes regions of high curvature by constraining the angle $\gamma_a$ between adjacent edges (as drawn in Figure \ref{Droplet})
\begin{equation}
    u_a^\text{angle} = \epsilon_\text{angle}\left(1 - \cos\left(\pi - \gamma_a \right) \right).
\end{equation}
To simulate an optical or magnetic trap pulling on the droplet, we add a harmonic potential term constraining the distance between the center of mass $\mathbf{r}_\text{com}$ of the points $\{\mathbf{R}_a \}_{a=1}^{N_{IB}}$ and the externally controlled trap position $\mathbf{R}_\text{ext}(t)$:
\begin{equation}
    u^\text{trap} = \frac{k_\text{trap}(t)}{2} \lvert\mathbf{R}_\text{ext}(t)- \mathbf{R}_\text{com}\rvert ^2. \label{eqUtrap}
\end{equation}
We let both $k_\text{trap}(t)$ and $\mathbf{R}_\text{ext}(t)$ depend on time so that the trap can be turned on and off as well as moved in space.  Figure \ref{Droplet} provides a schematic of the IB model.  

The droplet is assumed to be permeable, so that both the exterior and interior of the closed loop contain the fluid.  The interaction between the points $\mathbf{R}_a$ and the fluid is bi-directional.  In the boundary-to-fluid direction, the constitutive equation $U\left(\{\mathbf{R}_a \}_{a=1}^{N_{IB}} \right)$ produces forces $\mathbf{f}_a = -\nabla_{\mathbf{R}_a} U$ at the positions $\mathbf{R}_a$.  These forces are ``spread'' to the grid points of the LB domain through kernel-weighted sums over the IB points in the vicinity of the grid points.  Different choices of kernels are possible, and in this work we use
\begin{equation}
    K(\mathbf{r}) = \phi(r_x)\phi(r_y) / (\Delta x)^2,
\end{equation}
where 
\begin{equation}
    \phi(x) = \begin{cases}
    \frac{1}{4}\left(1 + \cos \left(\frac{\pi x}{2} \right) \right) & 0 \leq |x| \leq 2 \Delta x \\
    0 & 2\Delta x \leq |x|
    \end{cases}.
\end{equation}
The IB force $\mathbf{f}_{IB}(\mathbf{r})$ at lattice position $\mathbf{r}$ is then computed as 
\begin{equation}
    \mathbf{f}_{IB}(\mathbf{r}) = \sum_{a = 1}^{N_{IB}} \mathbf{f}_a K(\mathbf{R}_a - \mathbf{r}).
\end{equation}
For further details on this force-spreading procedure, see Refs.\ \citenum{kruger2017lattice},  \citenum{peskin2002immersed}, and  \citenum{kang2011comparative}.  The force $\mathbf{f}_{IB}$ enters as a contribution toward the external force $\mathbf{f}$ in Equation \ref{eqNS}, which is then accounted for in the LB algorithm following the approach introduced in Ref. \citenum{guo2002discrete}.  In the fluid-to-boundary direction, we assume that there is a no-slip condition between the boundary and the neighboring fluid, such that the boundary points $\mathbf{R}_a$ are simply carried along by the local fluid velocity $\mathbf{v}(\mathbf{R}_a)$.  The fluid velocities are only defined on the grid points, so $\mathbf{v}(\mathbf{R}_a)$ is interpolated at the point $\mathbf{R}_a$ using kernel-weighted sums over the lattice points in the vicinity of $\mathbf{R}_a$\cite{kruger2017lattice}:
\begin{equation}
\mathbf{v}(\mathbf{R}_a) = \sum_\mathbf{r} (\Delta x)^2 \mathbf{v}(\mathbf{r})K(\mathbf{R}_a - \mathbf{r}).
\end{equation}
For simplicity we assume here that the boundary motion is overdamped, so that $\partial_t \mathbf{R}_a = \mathbf{v}(\mathbf{R}_a)$.  In the timestep  $\Delta t$ the point $\mathbf{R}_a$ is then updated using a finite difference integration scheme in tandem with the LB iteration.  We use the predictor-corrector algorithm for all finite difference integration in this paper. It would be straightforward to relax both the no-slip condition, to account for finite friction between the boundary and fluid, and the overdamped condition, to account for inertia of the boundary\cite{buvsik2018dissipative}, but we do not pursue this here.

\section{Anisotropic elasticity tensor}\label{elasticityTensor}
The expression for the elasticity tensor $\widetilde{\mathbf{C}}(\mathbf{P})$ in the Mandel basis is given by the integral (cf.\ Equation \ref{eqCTens})
\begin{equation}
    \widetilde{C}_{\alpha \beta}(\theta_P, k; \widetilde{\mathbf{K}}^\text{x}) = \int_0^{2 \pi} g(\theta - \theta_P;k)\widetilde{K}_{\alpha \beta}(\theta) d\theta \label{eqCdef1}
\end{equation}
where $\theta_P$ and $k$ are one-to-one functions of $P$ and $\hat{\mathbf{P}}$, $g(\theta;k)$ is the von-Mises distribution, and for simplicity we have absorbed the prefactor $(\rho_\text{p} - \rho_\text{ref})^a$ into the definition of $\widetilde{\mathbf{K}}$.  The tensor $\widetilde{\mathbf{K}}(\theta)$ is given as a rotation of the constant tensor $\widetilde{\mathbf{K}}^\text{x}$ via Equation \ref{eqKMandel}, where $\widetilde{\mathbf{K}}^\text{x}$ is specified in Equation \ref{eqKmatrix}.  The rotation tensor $\widetilde{\mathbf{Q}}$ appearing in Equation \ref{eqKmatrix} is a function of the polar angle $\theta$ of $\hat{\mathbf{n}}$, given by 
\begin{equation}\label{qmatrix}
    \widetilde{\mathbf{Q}}(\theta) = 
    \begin{pmatrix}
    \cos ^2\theta & \sin ^2\theta & -2 \sin \theta \cos \theta \\
    \sin ^2\theta & \cos ^2\theta & 2 \sin \theta \cos \theta \\
    \sin \theta \cos \theta & -\sin \theta \cos \theta & \cos ^2\theta-\sin ^2\theta \\
    \end{pmatrix}.
\end{equation}
Evaluation of Equation \ref{eqCdef1} gives 
\begin{widetext}
\begin{align}
\widetilde{C}_{11} =\ & \bigg( \frac{\left(k \left(k^2+24\right) I_0(k)-8 \left(k^2+6\right) I_1(k)\right) \cos (4 \theta_P) (K^\text{x}_{1111}-2 K^\text{x}_{1122}-4
   K^\text{x}_{1212}+K^\text{x}_{2222})}{k^3} \nonumber \\
   &+I_0(k) (3 K^\text{x}_{1111}+2 K^\text{x}_{1122}+4 K^\text{x}_{1212}+3 K^\text{x}_{2222}) \nonumber \\
   &+4 I_2(k) (K^\text{x}_{1111}-K^\text{x}_{2222}) \cos (2
   \theta_P) \bigg) \bigg/ 8 I_0(k) \\
   \widetilde{C}_{12} =\ & \frac{1}{8} \bigg(-\frac{\left(k \left(k^2+24\right) I_0(k)-8 \left(k^2+6\right) I_1(k)\right) \cos (4 \theta_P) (K^\text{x}_{1111}-2 K^\text{x}_{1122}-4
   K^\text{x}_{1212}+K^\text{x}_{2222})}{k^3 I_0(k)} \nonumber \\
   & +K^\text{x}_{1111}+6 K^\text{x}_{1122}-4 K^\text{x}_{1212}+K^\text{x}_{2222}\bigg)  \\
   \widetilde{C}_{13} =\ &  \bigg( \frac{\pi  \left(k \left(k^2+24\right) I_0(k)-8 \left(k^2+6\right) I_1(k)\right) \sin (4 \theta_P) (K^\text{x}_{1111}-2 K^\text{x}_{1122}-4
   K^\text{x}_{1212}+K^\text{x}_{2222})}{k^3} \nonumber \\
   &+2 \pi  I_2(k) (K^\text{x}_{1111}-K^\text{x}_{2222}) \sin (2 \theta_P)\bigg) \bigg / 4 \pi  I_0(k)\\
   \widetilde{C}_{21} =\ & \widetilde{C}_{12} \\
   \widetilde{C}_{22} =\ & \widetilde{C}_{11} - \bigg(I_2(k) (K^\text{x}_{1111}-K^\text{x}_{2222}) \cos (2 \theta_P)\bigg) \bigg / I_0(k) \\
   \widetilde{C}_{23} =\ & -  \bigg( \frac{\pi  \left(k \left(k^2+24\right) I_0(k)-8 \left(k^2+6\right) I_1(k)\right) \sin (4 \theta_P) (K^\text{x}_{1111}-2 K^\text{x}_{1122}-4
   K^\text{x}_{1212}+K^\text{x}_{2222})}{k^3} \nonumber \\
   & + 2 \pi  I_2(k) (K^\text{x}_{2222}-K^\text{x}_{1111}) \sin (2 \theta_P) \bigg) \bigg /4 \pi  I_0(k) \\
   \widetilde{C}_{31} =\ & \bigg( \frac{\pi  \left(k \left(k^2+24\right) I_0(k)-8 \left(k^2+6\right) I_1(k)\right) \sin (4 \theta_P) (K^\text{x}_{1111}-2 K^\text{x}_{1122}-4
   K^\text{x}_{1212}+K^\text{x}_{2222})}{k^3} \nonumber \\
   &+2 \pi  I_2(k) (K^\text{x}_{1111}-K^\text{x}_{2222}) \sin (2 \theta_P)\bigg) \bigg /8 \pi  I_0(k) \\
   \widetilde{C}_{32} =\ & -\bigg( \frac{\pi  \left(k \left(k^2+24\right) I_0(k)-8 \left(k^2+6\right) I_1(k)\right) \sin (4 \theta_P) (K^\text{x}_{1111}-2 K^\text{x}_{1122}-4
   K^\text{x}_{1212}+K^\text{x}_{2222})}{k^3} \nonumber \\
   &+2 \pi  I_2(k) (K^\text{x}_{2222}-K^\text{x}_{1111}) \sin (2 \theta_P)\bigg) \bigg/8 \pi  I_0(k) \\
   \widetilde{C}_{33} =\ & \frac{1}{4} \bigg(-\frac{\left(k \left(k^2+24\right) I_0(k)-8 \left(k^2+6\right) I_1(k)\right) \cos (4 \theta_P) (K^\text{x}_{1111}-2 K^\text{x}_{1122}-4
   K^\text{x}_{1212}+K^\text{x}_{2222})}{k^3 I_0(k)} \nonumber \\
   &+K^\text{x}_{1111}-2 K^\text{x}_{1122}+4 K^\text{x}_{1212}+K^\text{x}_{2222}\bigg).
\end{align}
\end{widetext}

\section{Implementation details and parameterization}\label{Param}

The methods described in this paper were implemented in custom Julia\cite{bezanson2017julia} code.  Although the LB algorithm is amenable to parallelization we did not pursue this here.  In our implementation, $187,500$ simulation steps of the tensorial model on a grid of $62,500$ points runs in $\sim 20$ hours of wall time on a single CPU, while the scalar model runs in $\sim 4$ hours.  Much of the additional computation time for the tensorial model is due to evaluating the tensor $\mathbf{C}(\mathbf{P})$ at each point, as well integrating the dynamics for the field $\mathbf{P}(\mathbf{r})$.

The pulling protocol, defined by the curves $k_\text{trap}(t)$ and $\mathbf{R}_\text{ext}(t)$ which enter into Equation \ref{eqUtrap}, was observed to lead in some cases to instabilities in the simulation.  This occurred when the transitions between resting, pulling, and letting go were not sufficiently smooth, causing high-frequency fluctuations in the fluid.  To alleviate this, the pulling protocol is formed from successive sigmoid functions instead of step functions.  We use a width of $30$ timesteps for the sigmoid functions, and before pulling we first let the system equilibrate for $30$ timesteps. 

Several fields make up the system described in this paper, including $\rho$, $\mathbf{v}$, $\mathbf{R}_a$, $\boldsymbol{\sigma}^\text{p}$, and $\mathbf{P}$, and each requires specifying boundary conditions (BCs) at the edge of the simulation volume.  The fields $\rho$ and $\mathbf{v}$ are handled by the LB algorithm (see Figure \ref{Flowchart}) and typically have either periodic or Dirichlet BCs, though Neumann BCs are also possible.  In this paper we use Dirichlet BCs with zero wall velocity, implemented through the bounce-back method, which is described in detail in Ref.\ \citenum{kruger2017lattice}.  The IB points $\mathbf{R}_a$ are not prescribed BCs in this work; we simply ensure that these points do not cross the boundary during simulation.  The remaining fields $\boldsymbol{\sigma}^\text{p}$ and $\mathbf{P}$ are given Neumann BCs with zero derivative at the boundary, though it would be straightforward to use Dirichlet or Robin BCs for these fields as well \cite{adams2008interplay}.

We followed the general strategy for parameterizing the LB simulations described in Ref.\ \citenum{kruger2017lattice}.  We set the solvent viscosity to that of water, but, following standard practice with LB simulations, we set the solvent density to several orders of magnitude larger than the density of water \cite{tjhung2012spontaneous, cates2004simulating, wolff2012cytoplasmic, henrich2010ordering}.  This allows increasing the timestep (thereby accelerating simulations) while ensuring that the system is still in the low Reynolds number regime. To explore their effects, we varied the stress diffusion constant, droplet radius, and pulling time as described in the main text.  The default values of these parameters, as well as the viscoelastic parameters, were chosen to roughly correspond to pulling a small lipid droplet through the cytoplasm \cite{Xiee2115593119}.  The trap stiffness $k_\text{trap}$  was chosen to be roughly consistent with an an optical trap.  The droplet's constitutive parameters $k_\text{spring}$ and $\epsilon_\text{angle}$ were chosen to give appreciable resistance to deformation while remaining numerically stable.  The Beris-Edwards parameters are mostly based on previous works \cite{tjhung2012spontaneous}.  The parameter values we used for all simulations, unless otherwise specified, are given in Tables \ref{LatticeParams}-\ref{VEParams}.

\begin{table}
	\begin{center}
		\caption{Parameters related to the LB algorithm and system}\label{LatticeParams}
		\begin{tabular}{  l | l | l}
			\hline 
			\textbf{Parameter} & \textbf{Symbol} &  \textbf{Value}  \\ \hline 
			Lattice spacing & $\Delta x$  & $4 \times 10^{-8}$ m  \\
			Timestep & $\Delta t$  & $ 8 \times 10^{-6}$ s  \\
			Number of steps & $N_\text{steps}$  & $187,500$   \\
			Collision operator time & $\tau$ & 1.25 \\
			Solvent dynamic viscosity & $\eta_s$  & 0.001 Pa s \\
			Solvent density & $\rho_s$  & $2 \times 10^7$ kg/m$^3$   \\
			Lattice size & $N_x, \ N_y$  & $250, \ 250$  \\
            \hline
		\end{tabular}
	\end{center}
\end{table}

\begin{table}
	\begin{center}
		\caption{Parameters related to the IB droplet and pulling protocol}\label{IBParams}
		\begin{tabular}{  l | l | l}
			\hline 
			\textbf{Parameter} & \textbf{Symbol} &  \textbf{Value}  \\ \hline
			IB node distance & $l_\text{eq}$  & $2 \times 10^{-8}$ m  \\
			Droplet radius & $R$ & $5 \times 10^{-7}$ m \\
			Droplet spring stiffness & $k_\text{spring}$ & $10^{-5}$ N/m \\
			Droplet curvature stiffness & $\epsilon_\text{angle}$ & $10^{-20}$ N m \\
			Maximum trap stiffness & $k_\text{trap}$ & $10^{-6}$ N/m \\
			Droplet pulling distance & $d_\text{pull}$ & $4 \times 10^{-6}$ m \\
			Droplet pulling time & $T_\text{pull}$ & 0.2 s \\
            \hline
		\end{tabular}
	\end{center}
\end{table}

\begin{table}
	\begin{center}
		\caption{Parameters related to the polymer field dynamics}\label{BEParams}
		\begin{tabular}{  l | l | l}
			\hline 
			\textbf{Parameter} & \textbf{Symbol} &  \textbf{Value}  \\ \hline
			Flow alignment parameters & $\xi$  & 1.1  \\
			Polarization free energy coefficients & $\alpha, \ \beta$  & --0.9, 1.0  \\
	        Alignment stiffness & $\kappa$  & 0.001  \\
	        Rotational-diffusion constant & $\Gamma$  & 1.0  \\
            \hline
		\end{tabular}
	\end{center}
\end{table}

\begin{table}
	\begin{center}
		\caption{Parameters related to the viscoelastic stress$^a$}
		\begin{tabular}{  l | l | l}
			\hline 
			\textbf{Parameter} & \textbf{Symbol} &  \textbf{Value}  \\ \hline
			Scalar polymeric stiffness  & $C$ &  0.01 Pa  \\
			Polymeric viscosity  & $\eta_\text{p}$ &  0.1 Pa s  \\
			Stress diffusion constant  & $D_\text{p}$ &  $10^{-13}$ m$^2$/s  \\
			Stiffness tensor element & $K^\text{x}_{1111}$ &  0.01 Pa  \\
			Stiffness tensor element & $K^\text{x}_{1122}$ &  0.005 Pa  \\
			Stiffness tensor element & $K^\text{x}_{2222}$ &  0.005 Pa  \\
			Stiffness tensor element & $K^\text{x}_{1212}$ &  0.005 Pa  \\
            \hline
		\end{tabular}
		  
		  $^a$For simplicity, the prefactor $\left(\rho_\text{p} - \rho_\text{ref}\right)^a$ appearing in Equation \ref{eqCTens} has been absorbed into the values shown here.\label{VEParams}
	\end{center}
\end{table}

\clearpage
\bibliography{VELB.bib}
/%
\bibliographystyle{unsrt}

\end{document}